\documentclass[11pt]{article}   
\pdfoutput=1

\usepackage{amsthm,amsmath,amssymb,bbold,xcolor,sectsty,latexsym,mathtools,hyperref,mathrsfs,slashed,bm,epsfig}

\usepackage[textwidth=6.5in,top=1.2in,bottom=1.2in]{geometry}

\newcommand{\beq}{\begin{equation}}
\newcommand{\eeq}{\end{equation}}
\newcommand{\bea}{\begin{eqnarray}}
\newcommand{\eea}{\end{eqnarray}}
\newcommand{\amp}{&\!\!\!}

\newcommand{\phiamp}{\phi_0}
\newcommand{\Clm}{C_{lm}}
\newcommand{\fa}{f_a}
\newcommand{\ga}{g_{a\gamma}}
\newcommand{\tga}{\tilde{g}_{a\gamma}}
\newcommand{\ma}{m_\phi}
\newcommand{\ba}{\beta}
\newcommand{\om}{\omega_0}
\newcommand{\Phione}{\tilde\Phi_{1d}}
\newcommand{\La}{\text{l}}
\newcommand{\Ma}{\text{m}}
\newcommand{\gac}{g_{a\gamma,min}}

\begin{document}

\thispagestyle{empty}
\begin{titlepage}
\nopagebreak

\title{ \begin{center}\bf Dark Matter Axion Clump Resonance of Photons \end{center} }

\vfill
\author{Mark P.~Hertzberg$^{}$\footnote{mark.hertzberg@tufts.edu}, \,\,\, Enrico D.~Schiappacasse$^{}$\footnote{enrico.schiappacasse@tufts.edu}}

\date{\today}

\maketitle

\begin{center}
	\vspace{-0.7cm}
	{\it  $^{}$Institute of Cosmology, Department of Physics and Astronomy}\\
	{\it  Tufts University, Medford, MA 02155, USA}
	
\end{center}

\bigskip

\begin{abstract}
Recently there has been interest in the physical properties of dark matter axion condensates. Due to gravitational attraction and self-interactions, they can organize into spatial localized clumps, whose properties were examined by us in Refs.~\cite{Schiappacasse:2017ham,Hertzberg:2018lmt}. Since the axion condensate is coherently oscillating, it can conceivably lead to parametric resonance of photons, leading to exponential growth in photon occupancy number and subsequent radio wave emission. We show that while resonance always exists for spatially homogeneous condensates, its existence for a spatially localized clump condensate depends sensitively on the size of clump, strength of axion-photon coupling, and field amplitude. By decomposing the electromagnetic field into vector spherical harmonics, we are able to numerically compute the resonance from clumps for arbitrary parameters. We find that for spherically symmetric clumps, which are the true BEC ground states, the resonance is absent for conventional values of the QCD axion-photon coupling, but it is present for axions with moderately large couplings, or into hidden sector photons, or from scalar dark matter with repulsive interactions. We extend these results to non-spherically symmetric clumps, organized by finite angular momentum, and find that even QCD axion clumps with conventional couplings can undergo resonant decay for sufficiently large angular momentum. We discuss possible astrophysical consequences of these results, including the idea of a pile-up of clump masses and rapid  electromagnetic emission in the sky from mergers.
\end{abstract}

\end{titlepage}

\setcounter{page}{2}

\tableofcontents

\newpage

\section{Introduction}

It is essential to obtain clues of physics beyond the Standard Model. Since there is currently no evidence of new physics at colliders, including the LHC, we can turn to astrophysics and cosmology for possible clues. One definite example of new physics is the need for dark matter to comprise the bulk of the matter in the universe in order to be compatible with a range of astronomical observations, including CMB, large scale structure, galaxies and galactic halos, etc \cite{Peebles:2013hla}. There are a range of dark matter candidates, though currently there is no evidence for any of them. For example, the popular example of WIMP dark matter implies dimension 4 coupling to Standard Model particles via W or Z boson exchange. Yet there is currently no evidence for this coupling from a range of direct and indirect experiments, even though experiments have probed a significant part of parameter space where such WIMPs could easily have been detected by now.

Another popular dark matter candidate is the QCD axion \cite{Preskill:1982cy,Abbott:1982af,Dine:1982ah,Kim:2008hd}. It is a gauge singlet (pseudo)-scalar arising from the spontaneous breaking of a Peccei-Quinn (PQ) symmetry, introduced as a possible solution to the strong CP problem \cite{Peccei:1977hh,Weinberg:1977ma,Wilczek:1977pj}. The axion's (approximate) shift symmetry prevents it from having dimension 4 couplings to the Standard Model particles, making it very difficult to detect. On the other hand, it is expected to couple to the Standard Model via higher dimension operators. In particular, it should couple to photons via the dimension 5 operator $\Delta\mathcal{L}\sim \ga\,\phi\,{\bf E}\cdot{\bf B}$. Current constraints imply that the dimensionful coupling $\ga$ is quite small, though it may still have a value compatible with the QCD axion as a solution to the strong CP problem. For minimal axion models $\ga\sim\alpha/\fa$, where $\alpha$ is the fine structure constant and $\fa$ is the PQ breaking scale. In order for the QCD axion's abundance to not over-close the universe, the PQ scale should typically be $\fa\lesssim 10^{12}$\,GeV (although this bound can be relaxed depending on the details of inflation \cite{Hertzberg:2008wr}). Hence the coupling $\ga$ cannot be arbitrarily small. Such a coupling may lead to measurable effects. This axion-photon coupling is usually exploited to try to detect axions in ground based experiments, such as the ADMX experiment \cite{Asztalos:2009yp,Hoskins:2011iv}, in which dark matter axions move through a large magnetic field to produce a cavity photon. However, such effects have not currently been observed, though a detection is plausible in upcoming years.

It is therefore very important to explore possible consequences of this axion-photon coupling in other contexts; in particular, in astrophysical settings. In this paper we will investigate the possible consequences of this coupling on the behavior of small scale axion dark matter substructure. On small scales, axions can gravitationally thermalize leading to a type of Bose-Einstein condensate (BEC) \cite{Sikivie:2009qn,Erken:2011dz}. This condensate does not possess long-range order, as it is driven by attractive interactions: gravity and self-interactions $\lambda\,\phi^4$ with $\lambda<0$ \cite{Guth:2014hsa}. Instead the condensate is a spatially localized clump. The properties of these BEC clumps were analyzed in Ref.~\cite{Schiappacasse:2017ham} where we mapped out branches of stable and unstable solutions, finding that there is a maximum mass and a minimum radius for stable solutions. In Ref.~\cite{Hertzberg:2018lmt} we extended this analysis to BECs with angular momentum, finding that the maximum mass and minimum radius are both increased with angular momentum. 

Since the condensate is a coherently oscillating axion field, it can potentially lead to parametric resonance of the electromagnetic field from the axion-photon coupling, leading to an output of coherent radio waves. The phenomenon of parametric resonance is an important phenomenon in physics. One of the most striking consequences of this phenomenon is the drastic decay of the oscillating inflaton field into daughter fields at the first stage of reheating, known as preheating \cite{Kofman:1994rk, Shtanov:1995}. Other related work includes the study of photon propagation in a homogeneous cold axion field (and axion-like particles) in the presence of an external magnetic field ~\cite{Espriu:2011vj}. The subject of the present paper is the study of the possibility of parametric resonance of photons in the context of {\em localized} dark matter axion clumps; other related work includes Refs.~\cite{Tkachev:1986tr,Tkachev:1987cd,Tkachev:2014dpa}. 

To put this in context, we begin by considering homogeneous condensates (this may describe the behavior in the very early universe \cite{Yoshimura:1995gc,Yoshimura:1996fk}, although it quickly fragments \cite{Kitajima:2017peg}).  In the homogeneous case the equation of motion for the electromagnetic field becomes diagonal in $k$-space and organizes into a standard type of Matheiu equation.  This is amenable to the standard techniques of Floquet theory. In this homogeneous case the resonance is always present regardless of the strength of the coupling or the axion field amplitude (although the strength of the resonance is proportional to the product of these parameters).     

We then turn to our primary interest of possible resonance from the spatially localized clump condensate. In this case the equation of motion of the electromagnetic field couples all of its $k$-modes to one another since the axion clump breaks translation invariance. For spherically symmetric axion clumps, we decompose the electromagnetic field into vector spherical harmonics, and  focus on the channel $\{l,m\}=\{1,0\}$ by simplicity, leaving a more complete analysis for future work. For spatially wide axion clumps, which is the regime of most physical interest, we are able to entirely integrate out the angular dependence in the problem, leaving an effective 1-dimensional (radial) problem for the electromagnetic field's mode functions. This takes the form of an integro-differential equation in $k$-space, involving the convolution of the axion's 1-dimensional Fourier transform with the electromagnetic field's mode functions. We are able to readily solve this integro-differential equation numerically using a generalization of the Floquet theory to obtain the Floquet exponents. We find that unlike the homogeneous case, the presence of the resonance depends sensitively on the clump's field amplitude, axion-photon coupling, and spatial width. For typical values of the QCD axion-photon coupling $\ga\sim\alpha/f_a$ the resonance is shut-off for spherically symmetric clumps. However, for atypically large couplings $\ga\gtrsim 1/f_a$ the resonance is present for a range of clump masses and radii. Such a large coupling may be present in unconventional QCD axion models as well as coupling to hidden sector photons or from axion-like particles. We show that an excellent criteria for resonance is that the homogeneous Floquet exponent is greater than the escape rate of photons (roughly the inverse diameter of the clump), so as to make Bose-Einstein statistics effective. Finally, we consider resonance from non-spherical axion clumps organized by angular momentum. In this case we find that the resonance into photons is enhanced relative to the case of spherically symmetric clumps, as the radius of the clump is increased. For sufficiently large angular momentum we find that resonance is possible even for QCD axions with typical photon couplings.

We then discuss possible astrophysical consequences of our findings. In particular, we point out that large mass clumps could rapidly emit electromagnetic radiation, leading to a reduction in mass until eventually the resonance is shut-off and the clump mass becomes conserved. This predicts a build up of clump masses concentrated at a single mass determined purely in terms of fundamental constants. Furthermore, we point out that clump mergers could suddenly produce radiation in the universe today.

The outline of this paper is as follows: 
In Section \ref{AxionEssentials} we review some essentials of axion and electromagnetic field theory.
In Section \ref{HomogeneousCondensates} we discuss the simple case of resonance from a homogeneous condensate.
In Section \ref{SphericalCondensates} we examine in detail the case of resonance from spherically symmetric condensate clumps.
In Section \ref{Criteria} we discuss the general condition for resonance.
In Section \ref{Photonmass} we discuss the effect on resonance from the effective photon mass.
In Section \ref{AngularCondensates} we extend the above analysis to non-spherical condensate clumps organized by angular momentum. 
In Section \ref{Astrophysical} we discuss possible astrophysical consequences of our results.
Finally, in Appendix \ref{Appendix} we include some additional formulae.

\section{Axions and Photons}\label{AxionEssentials}

The basic dynamics of the QCD axion has been discussed in many papers. Here we only recap some essential features. The reader is referred to our previous papers \cite{Schiappacasse:2017ham,Hertzberg:2018lmt} for more details.

\subsection{Axion Field Theory}

The axion $\phi$ is a pseudo Goldstone boson associated with a spontaneously broken PQ symmetry. QCD instantons generate a small but non-zero potential $V$ for $\phi$. As a dark matter candidate, its occupancy number is expected to be huge, and so it is well described by classical field theory if a suitable ensemble averaging is performed \cite{Hertzberg:2016tal}. If we expand around the CP preserving vacuum $\phi=0$, the potential has only even powers of $\phi$ as follows
\beq
V(\phi)={1\over2}\ma^2\,\phi^2+{\lambda\over 4!}\,\phi^4+\ldots\,.
\eeq
Since we shall only be interested in the non-relativistic regime for axions, we shall focus on field configurations where $\phi$ is small and so we shall only need to track these leading terms. The specific values of the mass $\ma$ and quartic coupling $\lambda$ are model dependent. For the standard QCD axion, its mass is given in terms of the up and down quark masses, pion mass, pion decay constant, and PQ symmetry breaking scale $\fa$ as 
\beq
\ma^2={m_um_d\over(m_u+m_d)^2}{f_\pi^2 m_\pi^2\over f_a^2}\,,
\eeq
while the quartic coupling $\lambda$ is given by
\beq
\lambda = -\gamma {\ma^2\over f_a^2}<0,
\label{lambdachoice}\eeq
where $\gamma$ is an $\mathcal{O}(1)$ pre-factor. It is $\gamma=1$ in the standard dilute instanton gas approximation in which the axion potential is a simple cosine and $\gamma=1-3m_um_d/(m_u+m_d)^2\approx0.3$ in a more precise calculation \cite{diCortona:2015ldu}. (For repulsive self-interactions, we write $\lambda=+\gamma\,\ma^2/\fa^2>0$, keeping $\gamma$ positive.)

As we discussed in Refs.~\cite{Schiappacasse:2017ham,Hertzberg:2018lmt} the stable axion clump solutions are non-relativistic. These solutions are gravitationally bound clumps, or Bose stars (and related to miniclusters \cite{Kolb:1993zz}), and their behavior is influenced by the self-interaction from the above $\sim\lambda\,\phi^4$ term. In this non-relativistic regime we can re-write the real axion field $\phi$ in terms of a complex Schr\"odinger field $\psi$ as follows
\beq
\phi({\bf{x}},t)=\frac{1}{\sqrt{2\ma}}\left[e^{-i\ma t}\psi({\bf{x}},t)+e^{i\ma t}\psi^{*}({\bf{x}},t)\right]\,,
\label{phi}
\eeq
where $\psi$ is taken to be slowly varying in time. The axion field's oscillation frequency is approximately given by $\ma$, but there are small corrections provided by $\psi$. 

The dynamics of $\psi$ is given by the following standard non-relativistic Hamiltonian that respects the Galilean symmetry \cite{Guth:2014hsa,Schiappacasse:2017ham}
\beq
H_{nr} = H_{kin} + H_{int} + H_{grav}\,,
\label{HamTotal}
\eeq
where 
\bea
H_{kin} \amp=\amp {1\over2\ma}\int d^3x\, \nabla \psi^*\! \cdot\! \nabla \psi\,, \label{Hkin}\\
H_{int} \amp=\amp {\lambda\over 16\,\ma^2}\int d^3x \, \psi^{*2}\psi^2 \,,\label{Hint}\\
H_{grav} \amp=\amp -\frac{G\,\ma^2}{2}\int d^3x  \int d^3x' \frac{\psi^*({\bf{x}})\psi^*({\bf{x}}')\psi({\bf{x}})\psi({\bf{x}}')}{|{\bf{x}}-{\bf{x}}'|}\,,
\label{Hgrav}
\eea
are the kinetic energy $H_{kin}$, self-interaction energy $H_{int}$, and gravitational energy $H_{grav}$ terms, respectively. 
(See Ref.~\cite{Namjoo:2017nia} for an investigation into the leading relativistic corrections.)

In this non-relativistic regime, particle number changing processes are suppressed. Associated with this, the above Hamiltonian carries a global $U(1)$ symmetry $\psi\to\psi\,e^{i\theta}$ associated with a conserved particle number
\beq
N=\int d^3x\,\psi^*({\bf x})\psi({\bf x})\,.
\eeq
The axion condensate is specified by a fixed number of particles $N$. For the true BEC this corresponds to the state of minimum energy at fixed $N$, which is spherically symmetric. While another type of BEC corresponds to the state of minimum energy at fixed $N$ and angular momentum ${\bf L}$, which is non-spherical.

\subsection{Axion-Photon Interaction}

Even though in the non-relativistic limit of axions there is a conserved particle-number, there can still unavoidably be particle number changing processes from coupling to photons. The axion-photon decay channel runs through the chiral anomaly in which a fermion loop connects the axion with two photons. The Lagrangian density for the electromagnetic field is given by 
\beq
\mathcal{L}_{EM} = -\frac{1}{4}F_{\mu\nu}F^{\mu\nu}-\frac{\ga}{4}\phi \, F_{\mu\nu}\tilde{F}^{\mu\nu}\,,
\label{axionphotonlagrangiandensity}
\eeq
where $\tilde{F}^{\mu\nu}$ is the dual of the electromagnetic field strength tensor $\tilde{F}^{\mu\nu}={1\over2}\varepsilon^{\mu\nu\alpha\beta}F_{\alpha\beta}$ and $\ga$ is the axion-photon coupling constant. We can parameterize the axion-photon coupling $\ga$ in terms of a dimensionless coupling $\ba$ and the PQ scale as
\beq
\ga={\ba\over \fa}\,,
\eeq
where $\ba$ is model dependent. 

In conventional QCD axion models~\cite{Raffelt:1996,Kaplan:1985dv,Srednicki:1985xd} the coupling is given by
\beq
\ba=\frac{\alpha K}{2\,\pi}\,,
\label{axionphotoncoupling}
\eeq
where 
\beq
K =  \frac{\bar{E}}{\bar{N}}-\left(\frac{2}{3}\right)\frac{4+z+w}{1+z+w}\,.
\eeq
Here $\alpha$ is the fine structure constant, $z\equiv m_u/m_d$, and $w\equiv m_u/m_s$. The $\bar{E}$ and $\bar{N}$ quantities correspond to the electromagnetic and color anomalies related to the axion field. The ratio $\bar{E}/\bar{N}$ is present in models in which quarks and leptons carry both Peccei-Quinn and electrical charges. For instance, $\bar{E}/\bar{N} = 0$ ($K \approx -1.95$) in the standard KSVZ \cite{Kim:1979if,Shifman:1979if} model because the new exotic heavy quark fields which carry the Peccei-Quinn charge do not carry electromagnetic charge. By contrast, $\bar{E}/\bar{N} = 8/3$ ($K \approx +0.72$) in the DFSZ \cite{Dine:1981rt,Zhitnitsky:1980tq} or grand unified models because the given family of quarks and leptons carry both kind of charges. So in conventional QCD axion models $K$ is an $\mathcal{O}(1)$ number, leading to $\ba=\mathcal{O}(10^{-2})$. However, in unconventional axion models, $\ba$ can be larger than the above estimates and can even be $\mathcal{O}(1)$ in some exotic scenarios. In fact if we allow coupling to hidden sector photons, there is considerable freedom in the possible values of $\ba$ as it is essentially unconstrained by experiment. So in this paper we shall explore a range of values for $\ba$ including small to moderately large values. Finally, for ease of notation, we send $\ga\to|\ga|$ as only its magnitude is of significance here.

We work with a quantized four vector potential $\hat{A}^{\mu}=({\hat{A}_0,{\bf{\hat{A}}})}$ in a classical background given by the axion field.  We vary the above Lagrangian $\mathcal{L}_{EM}$ with respect to  $\hat{A}^{\mu}$ to obtain the Heisenberg equation of motion. For simplicity, we work in Coulomb gauge $\nabla\! \cdot {\bf{\hat{A}}}=0$. We self-consistently assume the axion field is slowly varying in space, which it must be within the non-relativistic approximation, and drop gradients of $\phi$. Then the equation of motion for the 2 propagating degrees of freedom of the photon ${\bf{\hat{A}}}$ can be written as
\beq
{\bf{\ddot{\hat{A}}}}-\nabla^2{\bf{\hat{A}}}+\ga \nabla \times(\partial_t\phi\, {\bf{\hat{A}}}) = 0\,.
\label{EMeom}
\eeq
In this expression we have moved the axion field $\partial_t\phi$ inside the spatial derivative $\nabla \times$ as we are neglecting gradients of the axion field. It is useful to write it in this form to make it manifest that we are in Coulomb gauge, as every term's divergence clearly vanishes. Then Fourier transforming to $k$-space, the final term becomes a convolution
\beq
{\bf{\ddot{\hat{A}}_k}}+k^2{\bf{\hat{A}_k}}+\ga \,i\,{\bf k}\times\!\int\! \frac{d^3{\bf{k'}}}{(2\pi)^3}\,\partial_t\phi_{{\bf{k}}-{\bf{k'}}}\, {\bf{\hat{A}_{k'}}} = 0\,.
\label{kspaceEMeom}
\eeq
Note that even though we are dropping gradient terms compared to time derivatives of $\phi$, i.e., $|\nabla\phi|\ll|\partial_t\phi|$ in the non-relativistic limit of axions, the spatial structure of $\phi$ can still be very important, as we explore in the upcoming sections.

\section{Homogeneous Condensates}\label{HomogeneousCondensates}

Since the axion field will undergo coherent oscillations in the classical field limit, we wish to explore possible parametric resonance into photons. As the simplest possible treatment of this behavior, let us begin by treating the axion field as homogeneous (also see Refs.~\cite{Yoshimura:1995gc,Yoshimura:1996fk}). In general such a configuration is unstable to collapse from gravity and attractive self-interactions, which ultimately try to drive a homogeneous condensate towards condensate clump solutions with short range correlations \cite{Guth:2014hsa}. This more realistic situation will be studied in detail in the next sections. Ignoring that for now, we treat the axion as homogeneous and oscillating periodically in its potential. For small field amplitudes the oscillations are approximately harmonic
\beq
\phi(t) = \phiamp\cos(\om\, t)\,,
\label{axionharmonic}
\eeq
where the amplitude of oscillation is $\phiamp$ and to an excellent approximation the frequency is $\om\approx \ma$.

Then the equation of motion for the electromagnetic modes decouple in $k$-space to become
\beq
{\ddot{\hat{A}}_{\bf k}}+k^2{\hat{A}_{\bf k}}-\ga\,\om\,\phiamp\sin(\om\,t)\, i\, {\bf{k}} \times {\hat{A}_{\bf k}} = 0\,.
\label{kspaceaxionhomogeneousphotonequation}
\eeq
We express the Fourier transform of the vector potential as
\beq
{\bf{\hat{A}_k}}(t) =\sum_{\lambda = \pm}\, \left[ \hat{a}_{\bf k, \lambda}\, {\boldsymbol{\epsilon}}_{{\bf k},\lambda}\, s_{\bf k}(t) +\hat{a}^{\dag}_{\bf k, \lambda}\, {\boldsymbol{\epsilon}}^{*}_{{\bf k},\lambda}\, s^{*}_{\bf k}(t) \right]\,,
\label{decompositionA}
\eeq
where ${\boldsymbol{\epsilon}}_{{\bf k},\lambda=\pm}$ and  $s_{\bf k}(t)$  correspond to vectors for circular polarization and the mode function, respectively, and $\hat{a}_{\bf k, \lambda}$ and $\hat{a}^{\dag}_{\bf k, \lambda}$ are annihilation and creation operators.  Noting that $i {\bf k} \times {\boldsymbol{ \epsilon}}_{{\bf k},\lambda}=k\,{\boldsymbol{ \epsilon}}_{{\bf k},\lambda}$, the polarizations decouple and the mode functions $s_{\bf k}(t)$ satisfy the classical equations of motion
\beq
{\ddot{s}_{\bf k}}+\left[k^2-\ga\,\om\,k\,\phiamp\sin(\om\,t)\right] {{s_{\bf{k}}}} = 0\,.
\label{kspaceaxionhomogeneousphotondecouplingequation}
\eeq
This is a special form of the Hill's equation, known as the Mathieu equation, which describes an oscillator with a periodic pump of frequency $\omega_k^2(t) = \omega_k^2(t+T)$, where $T=2 \pi/\om$ is the period of oscillations of the condensate. We rewrite  Eq.~\ref{kspaceaxionhomogeneousphotondecouplingequation} as
\beq
{{\ddot{s}_{\bf k}}}  + \omega_k^2(t){{s_{\bf k}}}=0\,,
\label{Hill}\eeq
where $\omega_k^2(t) = A + B \sin(\om\,t)$ with $A$ and $B$ coefficients given by 
\bea
A \amp= \amp k^2\,,\label{Ap}\\
B \amp = \amp -\ga\,\phiamp\, k\,\omega_0\,.
\label{Bp}
\eea
Note that the coupling between the mode function and the axion field depends on $k$. As we will analyze in detail in the following subsection, the periodicity of  $\omega_k(t)$ may lead to parametric resonance for modes with certain values of $k$.

\subsection{Small Amplitude Analysis}

The above Mathieu equation can be readily solved numerically; as we will do in the next subsection. For now we can provide a precise analytical result by operating in the small amplitude regime, where the pumping term is relatively small, correcting the free theory behavior of the electromagnetic field by a relatively small amount. It is well-known that in the parameter space of the Mathieu equation there is a band structure of unstable (resonant) and stable regions  (for standard textbooks about Mathieu equation and parametric resonance see Refs.~\cite{McLachlan1947, Erdelyi1955, LandauLifshitz}). While stable regions correspond to oscillatory solutions, unstable regions correspond to exponentially growing solutions. In general, solutions of Eq.~(\ref{Hill}) can be written  in the characteristic form
\beq
s_{\bf k}(t)= P_{\bf k}(t) e^{\mu_k t} + P_{\bf k}(-t) e^{-\mu_k t} \,,
\label{SolMathieuEquation}
\eeq
where the parameter $\mu_{k}$ is called the Floquet exponent and $P_{\bf k}(t)$ is a periodic function of time. When the Floquet exponent has a real part, the resonance phenomenon occurs. In the regime of small amplitude or weak coupling of $(k/\om) \gg (\ga \phiamp /2)$, we have a spectrum of narrow resonant bands equally spaced at $k^2 \approx (n/2)^2\, \om^2$ for $n=1,2,3,\dots$. The width of the resonance band and the Floquet exponent both decrease as increasing $n$. The exponential growing solutions for the mode function within the $n$-th resonant band, $s_{\bf k} \propto \exp(\mu_k^{(n)}\,t)$, are associated with an exponential growth of the occupation numbers, $n_{\bf k}(t) \propto \exp(2 \mu_k^{(n)}\,t)$, which can be seen as particle production.

In the small amplitude regime it is useful to expand the solution for the mode function, Eq.~(\ref{SolMathieuEquation}), as a harmonic expansion as follows
\beq
s_{\bf k}(t) = \sum_\omega e^{i\,\omega\, t}f_{\omega}(t)\,,
\label{harmonicexpansionmathieequationp}
\eeq
where the frequencies are summed over integer multiplies of half the natural $(\om/2)$ frequency and $-\infty  < \omega < \infty $. For small amplitudes, the functions $f_\omega(t)$ are {\em slowly varying}. Inserting this expansion into the Hill's equation and dropping the $\ddot{f}_\omega$ term, we obtain  
\beq 
4\,i\,\omega \dot{f}_{\omega}(t) + 2(A-\omega^2)f_{\omega}(t) - i\,B\left[ f_{\omega-\om}(t) - f_{\omega+\om}(t)  \right]  = 0\,.
\label{ODE1p}
\eeq
The lowest frequencies $\omega = \pm \om/2$ are dominant in the first instability band. Dropping all higher harmonics in Eq.~(\ref{ODE1p}), we obtain a coupled pair of differential equations for the lowest frequency modes as
\beq
{d\over dt}\!\left[ \begin{array}{c} f_{\frac{\om}{2}}(t) \\ f_{-\frac{\om}{2}}(t) \end{array} \right] 
= \frac{i}{m}\begin{bmatrix} A-\frac{\om^2}{4} & -i\frac{B}{2} \\ -i\frac{B}{2} & -A+\frac{\om^2}{4} \end{bmatrix} 
\left[ \begin{array}{c} f_{\frac{\om}{2}}(t) \\ f_{-\frac{\om}{2}}(t) \end{array} \right]\,.
\label{matrixequationp}
\eeq
Following standard matrix theory, the behavior of this system is determined by exponentials $\sim\exp\left(\pm \mu_k t\right)$, where the growth rate $\mu_k$ corresponds to the (positive) eigenvalue of the above matrix. Solving Eq.~(\ref{matrixequationp}), and using the expressions for $A$ and $B$, Eqs.~(\ref{Ap},\,\ref{Bp}), we obtain the growth rate
\beq
\mu_k = \sqrt{\frac{\ga^2\,k^2\,\phiamp^2}{4}-\frac{\left( k^2-\frac{\om^2}{4}  \right)^2}{\om^2}}\,.
\label{FEphoton}
\eeq
Since the resonance occurs for real Floquet exponents, edges of this instability band are given by values of $k$ at which the Floquet exponent becomes zero. The left and right hand edge are readily found to be 
\beq
k_{l/r,edge} = \sqrt{\frac{\om^2}{4}+\frac{\ga^2\,\om^2\,\phiamp^2}{16}}\pm \frac{\ga\,\om\, \phiamp}{4}\,,
\eeq
where the minus (plus) sign holds for the left (right) hand edge. The width of the first instability band is therefore proportional to the axion-photon coupling constant $\ga$ as 
\beq
\Delta k = k_{r,edge}-k_{l,edge}={\ga\,\om\, \phiamp \over2}\,.
\eeq
The center of the band $k^* = (\om/2)\sqrt{1+g_{a \gamma}^2\phiamp^2/2}$ can be approximated as $k^*\approx\ma/2$ for small amplitudes, which corresponds to an approximate effective frequency for the mode function, Eq.~(\ref{Hill}), given by $\omega_k^*(t) \approx k^*[ 1 - \, \ga\, \phiamp\, \sin(\ma t)]$. At the limit of weak coupling, the center of band $k^*$ agrees with the familiar perturbative decay process $\phi\to \gamma+\gamma$, which enforces this wavenumber by simple kinematics. However, the perturbative picture does not include effects from Bose-Einstein statistics when a state considerably increases its occupancy number and so the perturbative rate does not depend on the number of particles produced earlier. The standard perturbative decay, $\Gamma (\phi \rightarrow \gamma + \gamma) = (\ga^2 \ma^3)/(64 \pi)$ , is highly suppressed by the square of the axion-coupling constant. By contrast,  at small but finite amplitudes, Bose-Einstein statistics allows for exponential growth with the corresponding maximum Floquet exponent $\mu_H^*$ given by
\beq
\mu_H^* \approx {\ga\,\ma\,\phiamp\over4}\,,
\label{FloquetMax}
\eeq
where we have used an ``H" subscript to indicate that this result is only valid for the homogenous case.

Let us slightly exit the first instability band and enter to the adjacent  stable regions in the band structure by considering a value $k_{\pm} = k_{l/r,edge} \pm \delta k$, where the minus (plus) sign holds for $k_{l,edge}$ ($k_{r,edge}$). In this case the Floquet exponent in Eq.~(\ref{FEphoton}) becomes imaginary and the mode function, Eq.~(\ref{SolMathieuEquation}), merely oscillates with an approximate effective frequency given by $\omega_{k_{\pm}}(t) \approx \omega_{k_{l/r, edge}}(t) \pm \delta k$, where the minus (plus) sign holds for $k_{l,edge}$ ($k_{r,edge}$).  Note that the condition for small amplitude always ensures the reality of the effective frequency. 

In principle, we could continue our analysis and focus on the second instability band, in which the leading harmonics are $\omega =\pm \om$, which is connected to $\phi+\phi\to\gamma+\gamma$. However the dominant contribution  comes from the first band as the higher order processes are suppressed at small amplitude. Also, let us comment on backreaction. As the electromagnetic field energy increases due to resonance, it draws energy away from the axion field. This is a higher order process and is beyond the scope of the present paper. However the leading order resonant behavior suffices to capture the basic physical process.

\subsection{Numerical Analysis}\label{NumericalHom}

We can also solve Eq.~(\ref{Hill}) exactly using Floquet theory. Following the standard Floquet method, we consider the following orthogonal set of initial conditions for the pair $(s_{\bf k},\,{\dot s}_{\bf k})$
\beq
\left(\begin{array}{cc}
s_{\bf k} \\
\dot{s}_{\bf k}
\end{array}\right)_{initial}
\to
\mathbb{1}_{2}=
\left(\begin{array}{cc}
1 & 0 \\
0 & 1
\end{array}
\right)\,.
\eeq
Then numerically evolve the system using Eq.~(\ref{Hill}) through one period $T=2\pi/\om$ of the axion pump field, which maps this initial $2\times2$ identity matrix $\mathbb{1}_2$ to a new $2\times2$ matrix $\mathbb{M}_2$. Then through $P$ periods, the system will evolve to $\mathbb{M}_2^P$. 
Hence the behavior is dictated by the 2 eigenvalues $\chi_k$ of the matrix $\mathbb{M}_2$, with Floquet exponents given by $\mu_k=\ln\chi_k/T$.

\begin{figure}[t]
\centering
\includegraphics[scale=0.38]{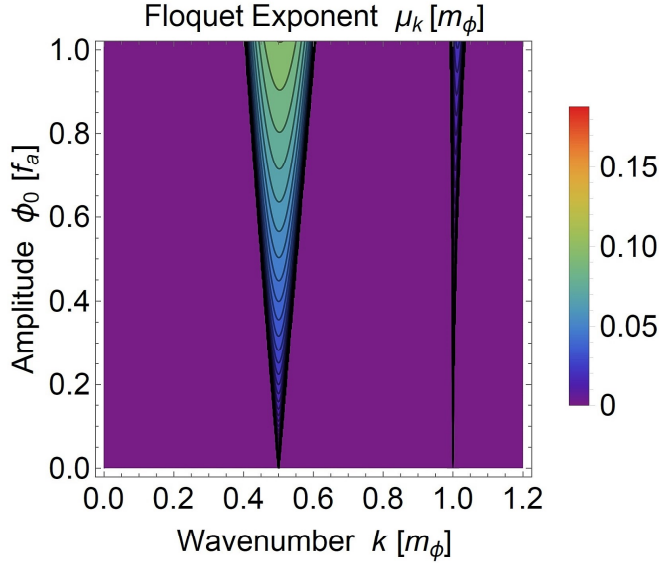}\!\!\!\!\!\!\!\!\!\!\!\!\!
\caption{Contour plot of the real part of Floquet exponent $\mu_k$, describing parametric resonance of photons from a homogeneous condensate, as a function of wavenumber $k$ and physical amplitude $\phiamp$. We plot $\phiamp$ in units of $\fa$ and $k$ \& $\mu_k$ in units of $\ma$. We have set $\ga=0.4/\fa$ to illustrate the behavior, although in conventional QCD axions $\ga=\mathcal{O}(10^{-2})/f_a$ (see Eq.~\ref{axionphotoncoupling}) which would give narrower resonance bands.}
\label{FloquetHomogeneous}
\end{figure}

The results are plotted in Figure~\ref{FloquetHomogeneous}. This shows the contour plot of the real part of the Floquet exponent as a function of wavenumber $k$ and physical amplitude $\phiamp$. We have set $\ga = 0.4\sqrt{\gamma}/\fa$ for illustrative purposes, though more conventional values for the QCD axion are $\ga=\mathcal{O}(10^{-2})/\fa$. The first instability band starts at $k = \ma/2$, and its structure agrees well with teh above analytical approximation in Eq.~(\ref{FEphoton}). Higher instability bands start at integer multiples of $k = \ma/2$. 

\section{Spherically Symmetric Clump Condensates}\label{SphericalCondensates}

We now turn to the physically important case of condensates that are spatially localized rather than merely homogenous as studied in the previous section. Gravity will inevitably cause a homogeneous condensate to fragment into an inhomogeneous field configuration; locally this leads to the formation of BEC clumps. When gravity is dominant these are so-called Bose stars. The self-interactions $\sim\lambda\,\phi^4$ can also play an important role too, and generically we shall refer to the solutions as ``clumps". When $\lambda<0$, as is expected for the QCD axion, this additional attraction can make the clump collapse under some conditions. There is an additional solution branch for highly dense clumps, known an ``axitons" \cite{Kolb:1993hw}, which are very short lived as they radiate semi-relativistic axions, and will not be our focus here. (See Ref.~\cite{Visinelli:2017ooc} for a recent investigation into their properties.)

\subsection{Clump Profile}\label{ClumpProfile}

The true BEC ground state is guaranteed to be spherically symmetric and will be discussed in this section. (See the next section for non-spherical BEC's with angular momentum.) These spherically symmetric clumps were studied in detail by us recently in Ref.~\cite{Schiappacasse:2017ham} (other work includes Refs.~\cite{Chavanis:2011zi,Chavanis:2011zm}) and we recap their important features briefly here. 

The non-relativistic Schr\"odinger field $\psi$ can be written as
\beq
\psi(r,t) = \Psi(r)\,e^{-i\,\mu\,t}\,,
\eeq
where $\Psi(r)$ describes the radial profile of the clump and $\mu$ describes the correction to the frequency. In Ref.~\cite{Schiappacasse:2017ham} we determined $\Psi(r)$ exactly numerically. We also discussed several approximate forms for $\Psi$. One form that was found to be highly accurate was to take $\Psi(r)$ to be a sech function
\beq
\Psi(r) =\sqrt{3\,N\over\pi^3\, R^3}\, \mbox{sech}(r/R)\,\,\,\,\,(\mbox{sech ansatz})\,,
\label{SechAnsatz}\eeq
where $R$ sets the effective radius of the solution and plays the role of a variational parameter. By inserting this into Eq.~(\ref{HamTotal}) and integrating, one finds the following form for the energy
\beq
H(R) = a\frac{N}{2\ma\,R^2} - b\frac{G\,\ma^2\,N^2}{R} +c\frac{\lambda\,N^2}{\ma^2\,R^3 }\,,
\label{HamSech}
\eeq
where in this sech ansatz we have coeficients
\beq
a={12+\pi^2\over6\pi^2},\,\,\, b={6(12\,\zeta(3)-\pi^2)\over\pi^4},\,\,\, c={\pi^2-6\over 8\pi^5}\,.
\eeq

\begin{figure}[t]
\centering
\includegraphics[scale=0.4]{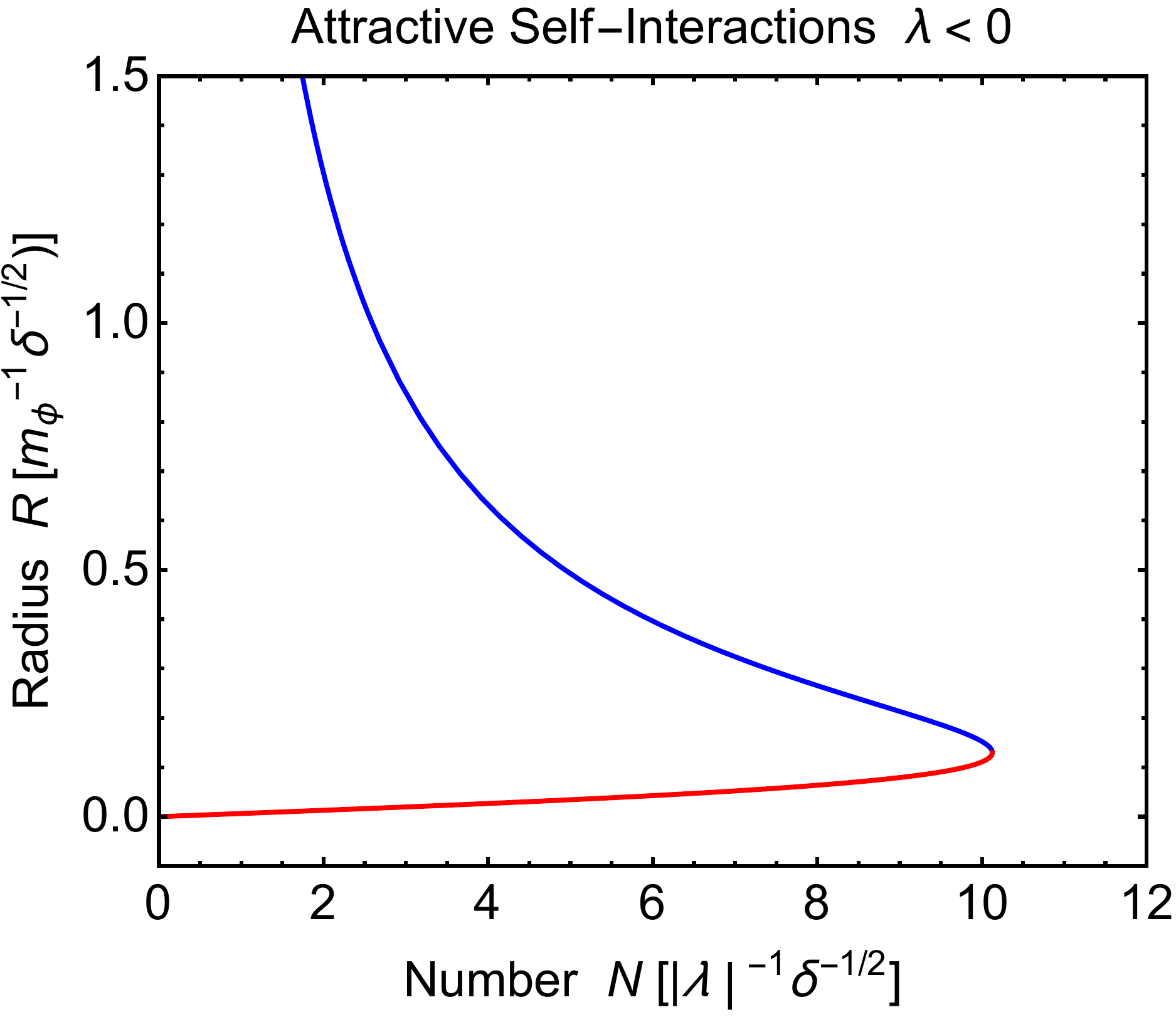}\,\,\,\,\,
\includegraphics[scale=0.4]{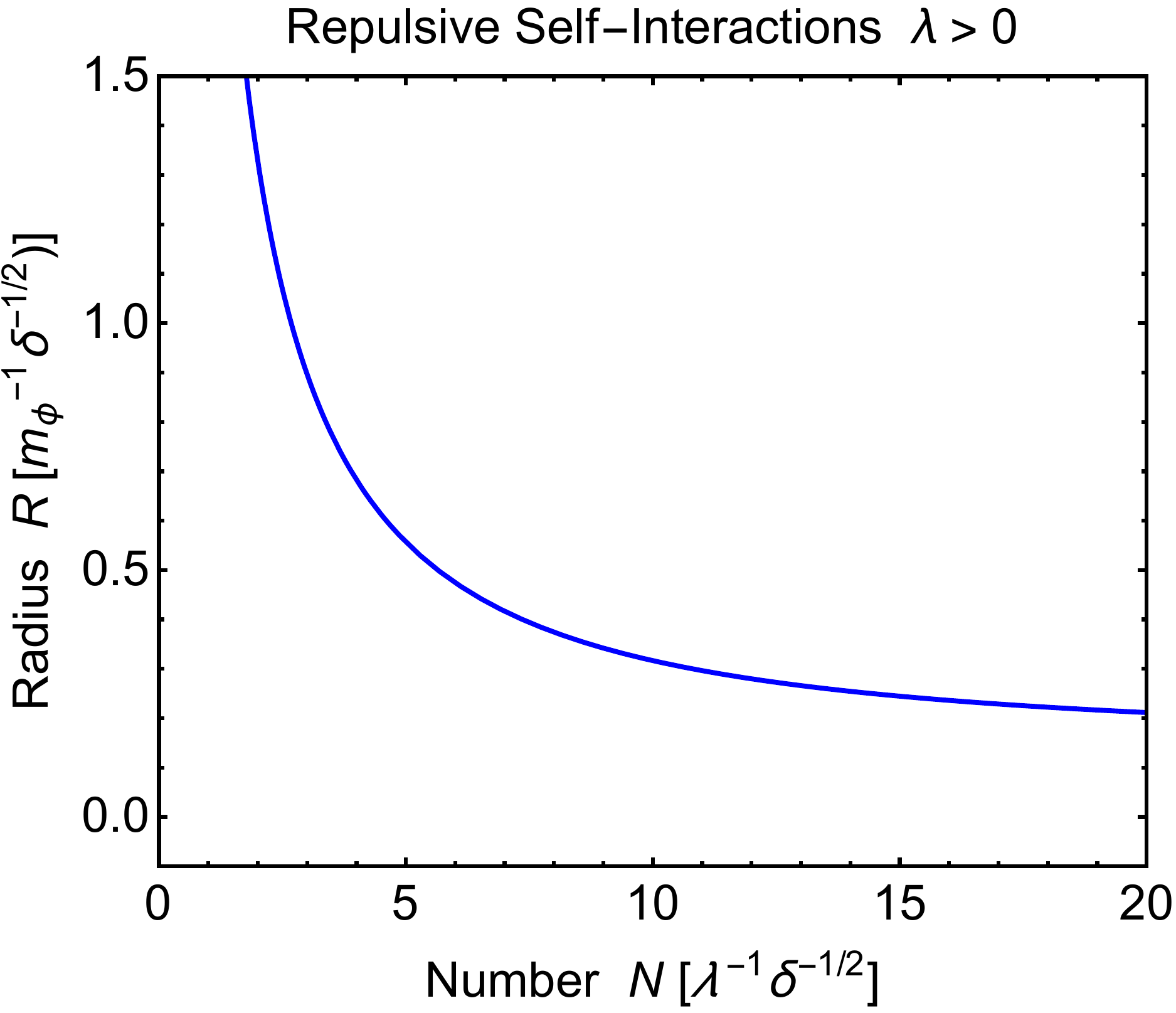}
\caption{Clump radius $R$ as a function of particle number $N$ for spherically symmetric clumps in the sech profile approximation (which is known to be very accurate from Ref.~\cite{Schiappacasse:2017ham}). We plot $R$ in units of $\ma^{-1}\delta^{-1/2}$ and $N$ in units of $|\lambda|^{-1}\delta^{-1/2}$. Left panel: Attractive self-interactions $\lambda<0$, leading to an upper blue stable branch and a lower red unstable branch. Right panel: Repulsive self-interactions $\lambda>0$, leading to an upper blue stable branch only.}
\label{FigureRadiusNumber}
\end{figure}

By extremizing the Hamiltonian with respect to $R$ at fixed $N$ one finds equilibrium solutions. For  attractive interactions $\lambda<0$ there are 2 branches of solutions, which are plotted in the left panel of Fig.~\ref{FigureRadiusNumber}; the upper blue branch are stable solutions and will be the focus of our investigation here, while the lower red branch are unstable solutions and will be ignored here. In this case there is a maximum number in the clump given by
\beq
N_{max} =  {a\over\sqrt{3\,b\,c}}{1\over|\lambda|\sqrt{\delta}} \approx {10.12\over |\lambda|\sqrt{\delta}}\,,
\label{Nmax}\eeq
and a minimum radius on the stable branch of $R_{min}=\sqrt{3\,c/b}/(\ma\sqrt{\delta})$, where the dimensionless quantity $\delta$ is defined as
\beq
\delta\equiv{G\,\ma^2\over|\lambda|}={G\,f_a^2\over\gamma}\,.
\eeq
In fact $\delta$ sets the characteristic squared-speed of particles in the clump and is very small for usual parameters of the QCD axion in which $\fa\ll1/\sqrt{G}$ is expected. For repulsive interactions $\lambda>0$ there is only 1 branch of solution, which is plotted in the right panel of Fig.~\ref{FigureRadiusNumber}; it is stable and will also be studied here. In this case there is no maximum number.

The corresponding relativistic field $\phi$ is given by exact harmonic oscillations in this non-relativistic regime of the form
\beq
\phi(r,t) = \Phi(r)\cos(\om\,t)\,,
\eeq
where the radial profile is
\beq
\Phi(r)=\sqrt{2\over\ma}\,\Psi(r)\,,
\eeq
and the oscillation frequency is $\om=\ma+\mu\approx\ma$. This can possibly lead to resonance of the electromagnetic field as we now examine.

\subsection{Vector Spherical Harmonic Decomposition}\label{Vector}

In order to examine possible resonance, let us return to Eq.~(\ref{EMeom}) for the equation of motion of the electromagnetic field. Since the axion field is spherically symmetric $\phi=\phi(r,t)$, but radially dependent, the usual 3-dimensional Fourier transform to Eq.~(\ref{kspaceEMeom}) is not the most efficient way to proceed as the vector structure remains complicated. Instead it is convenient to exploit the spherically symmetry of the axion field and re-organize the system into an effective 1-dimensional problem.

However, we cannot merely assume that the quantized vector potential ${\bf \hat{A}}$ is spherically symmetric as this would only allow for longitudinal modes. These are forbidden for massless photons and are manifestly removed in Coulomb gauge $\nabla\cdot{\bf \hat{A}}=0$. Instead, even though we will be interested in spherical waves, we need to allow for angular dependence in the vector potential. To proceed, we perform a vector spherical harmonic decomposition of ${\bf \hat{A}}$ as follows
\beq
{\bf \hat{A}}({\bf x},t) = \int\!{d^3k\over(2\pi)^3}\sum_{lm}\left[\hat{a}(k)\,v_{\La\Ma}(k,t){\bf M}_{\La\Ma}(k,{\bf x}) -\hat{b}(k)\, w_{\La\Ma}(k,t){\bf N}_{\La\Ma}(k,{\bf x}) + h.c.\right],\,
\eeq
where the ``vector spherical harmonics" ${\bf M}_{\La\Ma}$ and ${\bf N}_{\La\Ma}$ are defined in terms of the scalar spherical harmonics $Y_{\La\Ma}=Y_{\La\Ma}(\theta,\varphi)$ and spherical Bessel functions $j_\La$ as
\bea
{\bf M}_{\La\Ma}(k,{\bf x}) & = & {i\,j_\La(k\,r)\over\sqrt{\La(\La+1)}}\left[{i\,\Ma\over\sin\theta}Y_{\La\Ma}\hat\theta-{\partial Y_{\La\Ma}\over\partial\theta}\hat\varphi\right]\,,\label{Mvsh}\\
{\bf N}_{\La\Ma}(k,{\bf x}) & = & {i\over k}\nabla\times{\bf M}_{\La\Ma}(k,{\bf x})\,.\label{Nvsh}
\eea
where $r=|{\bf x}|$ is radius, $\theta$ is polar angle, and $\varphi$ is azimuthal angle. All the above quantities are taken to be functions of only the magnitude of the wavevector $k=|{\bf k}|$ since we are interested in spherical waves. The two scalar functions $v_{\La\Ma}(k,t)$ and $w_{\La\Ma}(k,t)$ are the electromagnetic mode functions. Naturally there are two independent mode functions as there are two polarizations of electromagnetic waves. Note that in addition to Eq.~(\ref{Nvsh}) we also have the inverse relation ${\bf M}_{\La\Ma}=-i\,\nabla\times {\bf N}_{\La\Ma}/k$; together this guarantees that $\nabla\!\cdot\!{\bf M}_{\La\Ma}=\nabla\!\cdot\!{\bf N}_{\La\Ma}=0$ as required by Coulomb gauge.

By inserting this decomposition into Eq.~(\ref{EMeom}) and again neglecting gradients of the axion field, the equations of motion for the mode functions $v_{\La\Ma}$ and $w_{\La\Ma}$ are  
\bea
&&\int\!{d^3k\over(2\pi)^3}\sum_{lm}\Big{[}\left(\ddot v_{\La\Ma}+k^2 v_{\La\Ma}-i\,k\,\ga\,\partial_t\phi\,w_{\La\Ma}\right){\bf M}_{\La\Ma}\nonumber\\
&&\,\,\,\,\,\,\,\,\,\,\,\,\,\,\,\,\,\,\,\,\,\,\,\,\,\,\,\,\,\,-\left(\ddot w_{\La\Ma}+k^2 w_{\La\Ma}+i\,k\,\ga\,\partial_t\phi\,v_{\La\Ma}\right){\bf N}_{\La\Ma} \Big{]} = 0\,,\,
\label{VECeom}\eea
which in principle can be solved numerically. However for an arbitrary sum over $\{\La,\Ma\}$, this is quite complicated.

\subsection{Resonance Channel}\label{Dominant}

We expect that several values of $\{l,m\}$ contribute in the classical equation of motion for the mode functions of the vector potential, Eq.~(\ref{VECeom}). Here we focus on the channel $\{l=1,m=0\}$ by simplicity and we leave a more complete analysis for future work.

By focussing then on $v_{10}$ and $w_{10}$, with all other mode functions set to zero, we can explore this channel for resonance. We can readily write out the individual vector components $(\hat{r}, \hat\theta, \hat\varphi)$ of Eq.~(\ref{VECeom}) for $\La=1$ and $\Ma=0$ giving
\bea
&& \int\!{d^3k\over(2\pi)^3}\left[\ddot{w}_{10}+k^2\,w_{10}+i\,k\,\ga\,\partial_t\phi\,v_{10}\right] {j_1(k\,r)\over k\,r}Y_{10} = 0\,\,\,\,\,\,(\hat r\,\,\,\mbox{component})\,,\label{rcomp}\\
&& \int\!{d^3k\over(2\pi)^3}\left[\ddot{v}_{10}+k^2\,v_{10}-i\,k\,\ga\,\partial_t\phi\,w_{10}\right]j_1(k\,r)Y_{11}\,e^{-i\varphi} = 0\,\,\,\,\,\,(\hat\varphi\,\,\,\mbox{component})\label{varphicomp}\,.
\eea
There is a similar equation for the $\hat{\theta}$ component, but it is automatically satisfied once the $\hat{r}$ and $\hat{\varphi}$ equations are satisfied. This is because we are in Coulomb gauge $\nabla\!\cdot\!{\bf \hat{A}}=0$, which reduces the system to only two independent equations.

Consider the radial component Eq.~(\ref{rcomp}). We substitute $\phi=\Phi(r)\cos(\om\,t)$ and multiply the expression by $Y_{10}^*\, j_{1}(k'\,r)\,k'\,r$ and integrate the whole expression over space $\int \! d^3x$. This gives
\beq
\ddot{w}_{10}(k',t)+k'^2\,w_{10}(k',t)-{2\,i\over\pi}\ga\,\om\,k'\sin(\om\,t)\!\int \!dk\,k^2\,v_{10}(k,t) \!\int \!dr\,r^2\,\Phi(r)\,j_1(k\,r)\,j_1(k'\,r) = 0\,,
\label{weqn}\eeq
where we have used the orthogonality property of the spherical Bessel functions
\beq
\int\! dr\,r^2\,j_1(k\,r)\,j_1(k'\,r)={\pi\,\delta(k-k')\over2\,k^2}\,,
\eeq
to simplify the first two terms in Eq.~(\ref{weqn}). 

The third term in Eq.~(\ref{weqn}) appears complicated, however we can simplify its form by the following method. Even though our system is actually 3-dimensional, the spherically symmetry of the axion field means that we can represent the axion's spatial profile $\Phi(r)$ by a 1-dimensional (real) Fourier transform $\Phione(k)$
\beq
\Phi(r) = \int{d\tilde k\over2\pi}\, \cos(\tilde{k}\,r)\,\Phione(\tilde k)\,.
\eeq
We re-express the product of the spherical Bessel functions by the identity
\beq
j_1(k\,r)\,j_1(k'\,r) = {1\over 4k^2k'^2 r}\int_{|k-k'|}^{k+k'}dk''\sin(k''\,r)(k^2+k'^2-k''^2)\,.
\eeq
By inserting this into the third term in Eq.~(\ref{weqn}) we can now carry out the $\int \!dr$ integral using
\beq
\int dr\,r\,\sin(k'' r)\,\cos(\tilde{k}\,r) = -{\pi\over2}{\partial\over\partial k''}\left[\delta(\tilde{k}+k'')+\delta(\tilde k-k'')\right]\,,
\eeq
and then performing the $\int\! d\tilde{k}$ integral to obtain
\beq
\int \!dr\,r^2\,\Phi(r)\,j_1(k\,r)\,j_1(k'\,r) = -{1\over 8k^2k'^2} \int_{|k-k'|}^{k+k'}dk''(k^2+k'^2-k''^2){\partial\over\partial k''}\Phione(k'')\,.
\label{Int}\eeq

In order to evaluate this integral, we recall that we are interested in axion field configurations that are slowly varying in space. So their Fourier transforms are concentrated around $k''\approx 0$. For example, for the sech ansatz of Eq.~(\ref{SechAnsatz}) the 1-dimensional Fourier transform is also a sech function
\beq
\Phione(k) = \sqrt{3\,N\over\pi\,R} \,\mbox{sech}\left(\pi\,k\,R\over 2\right) \,\,\,\,\,(\mbox{sech ansatz})\,.
\eeq
For large radius $R$ this is concentrated near small values $k\sim 1/R\ll\ma$, which is required for the non-relativistic approximation of the axion to be valid. 

In this large $R$ regime (and allowing for general profiles, rather than only the sech) the integral in Eq.~(\ref{Int}) can be evaluated by taking $k^2+k'^2-k''^2\approx k^2+k'^2$ in the integrand giving
\beq
\int \!dr\,r^2\,\Phi(r)\,j_1(k\,r)\,j_1(k'\,r) \approx {k^2+k'^2\over 8k^2k'^2}\left[\Phione(k-k')-\Phione(k+k')\right]\,.
\label{Phionesum}\eeq
Now the resonance occurs when the wavenumbers satisfy $k\approx k'\approx\om/2\approx\ma/2$. In this regime we can ignore the second term in Eq.~(\ref{Phionesum}) as it is exponentially suppressed for $R\gg 1/\ma$. Inserting this result into Eq.~(\ref{weqn}) (and interchanging $k$ with $k'$) we obtain
\beq
\ddot{w}_{10}(k,t)+k^2\,w_{10}(k,t)-i\,\ga\,\om\,k\sin(\om\,t)\!\int \!{dk'\over2\pi}\,v_{10}(k')\,\Phione(k-k') = 0\,.
\label{weqnbetter}\eeq
A similar line of reasoning goes through for the angular component Eq.~(\ref{varphicomp}), leading to
\beq
\ddot{v}_{10}(k,t)+k^2\,v_{10}(k,t)+i\,\ga\,\om\,k\sin(\om\,t)\!\int \!{dk'\over2\pi}\,w_{10}(k')\,\Phione(k-k') = 0\,.
\label{veqnbetter}\eeq
This pair of coupled equations for the mode functions $v_{10}$ and $w_{10}$ can be studied numerically. A self-consistent resonant solution is obeyed by
\beq
w_{10}(k,t)=\pm i\,v_{10}(k,t)\,,
\eeq
which reduces the system to a single scalar differential equation and is effectively 1-dimensional. This is much more tractable than the general 3-dimensional form mentioned earlier in Eq.~(\ref{kspaceEMeom}).

\subsection{Numerical Method}

In order to compute the resonance structure numerically, we need to generalize the Floquet theory used earlier in Section \ref{NumericalHom}. Since all the $k$-modes are now coupled to each other, we discretize our 1-dimensional $k$-space as
\beq
k = {2\,\pi\,n \over L}\,,\,\,\,\,\, n\in\mathbb{Z}\,,
\eeq
where $L$ is the size of the integration box which should be taken to be much larger than the size of a clump, i.e., $L\gg R$. This means we now have a set of coupled oscillators with a sinusoidal time dependent coupling. 

Recall that in the homogenous case, as discussed in Section \ref{HomogeneousCondensates}, the resonance occurred for wavenumbers $k\approx\om/2\approx\ma/2$. Since we are considering wide axion clumps, this general idea will persist with $k\sim\om/2\approx\ma/2$ in the vicinity of the resonance. So we take a finite set of wavenumbers, say $K$ values, that surround $\ma/2$. To perform the generalized Floquet analysis, we consider a column vector of length $K$ of mode functions $\vec{v}_{10}$ whose elements corresponds to each $k$-value. We then form a $2K\times2K$ matrix of initial conditions which spans the complete space of solutions for the mode functions as
\beq
\left(\begin{array}{cc}
\vec{v}_{10} \\
\dot{\vec v}_{10}
\end{array}\right)_{initial}
\to
\mathbb{1}_{2K}=
\left(\begin{array}{cc}
\mathbb{1}_K & \mathbb{0}_K \\
\mathbb{0}_K & \mathbb{1}_K
\end{array}
\right)\,.
\eeq
We then numerically evolve this set of coupled equations through one period $T=2\pi/\om$ of the axion oscillation to obtain a new matrix $\mathbb{M}_{2K}$. Then, as we mentioned earlier in the homogenous case, the behavior is controlled by the $2K$ eigenvalues $\chi$ of the matrix $\mathbb{M}_{2K}$, with Floquet exponents $\mu=\ln\chi/T$. These are not labelled by wavenumber anymore, as Fourier modes do not diagonalize the system. Nevertheless we can report on the maximum Floquet exponent $\mu^*$ which will dominate at late times.

\subsection{Numerical Results}

We have numerically determined the maximum Floquet exponent $\mu^*$ for this system for various choices of axion-photon coupling $\ga$ and for various parameters of the axion clump specified by its radius $R$ and particle number $N$. Operating in the sech approximation; physical clumps have a simple relationship between radius and number that we recapitulated earlier in Section \ref{ClumpProfile} and summarized in Fig.~\ref{FigureRadiusNumber}. We will focus on the stable blue branch here. This allows us to eliminate radius $R$ in favor of particle number $N$.

\begin{figure}[t]
\centering
\includegraphics[scale=0.31]{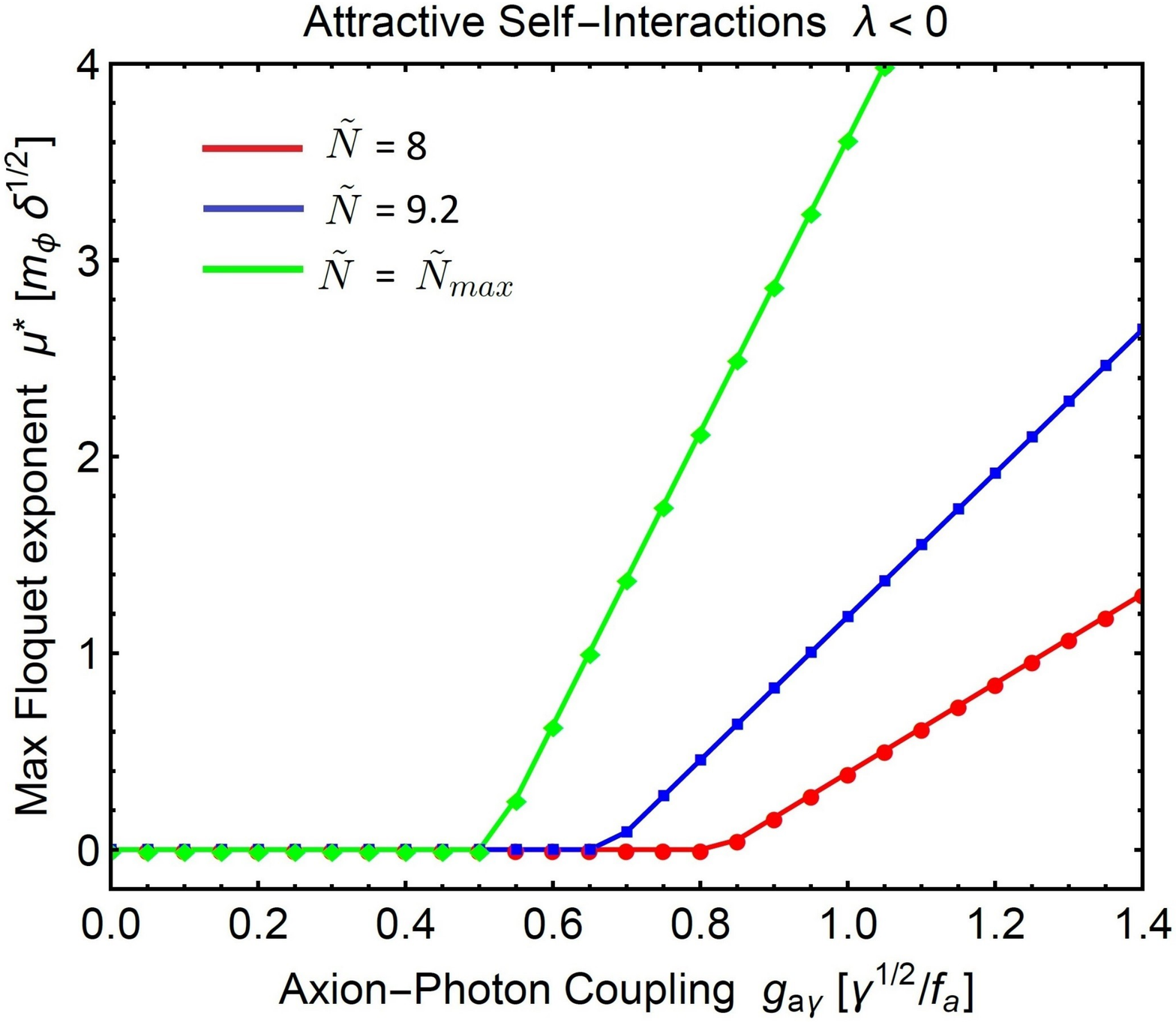}\!\!\!\!\!\!\!\!\!\!\!\!\!\!\!
\includegraphics[scale=0.31]{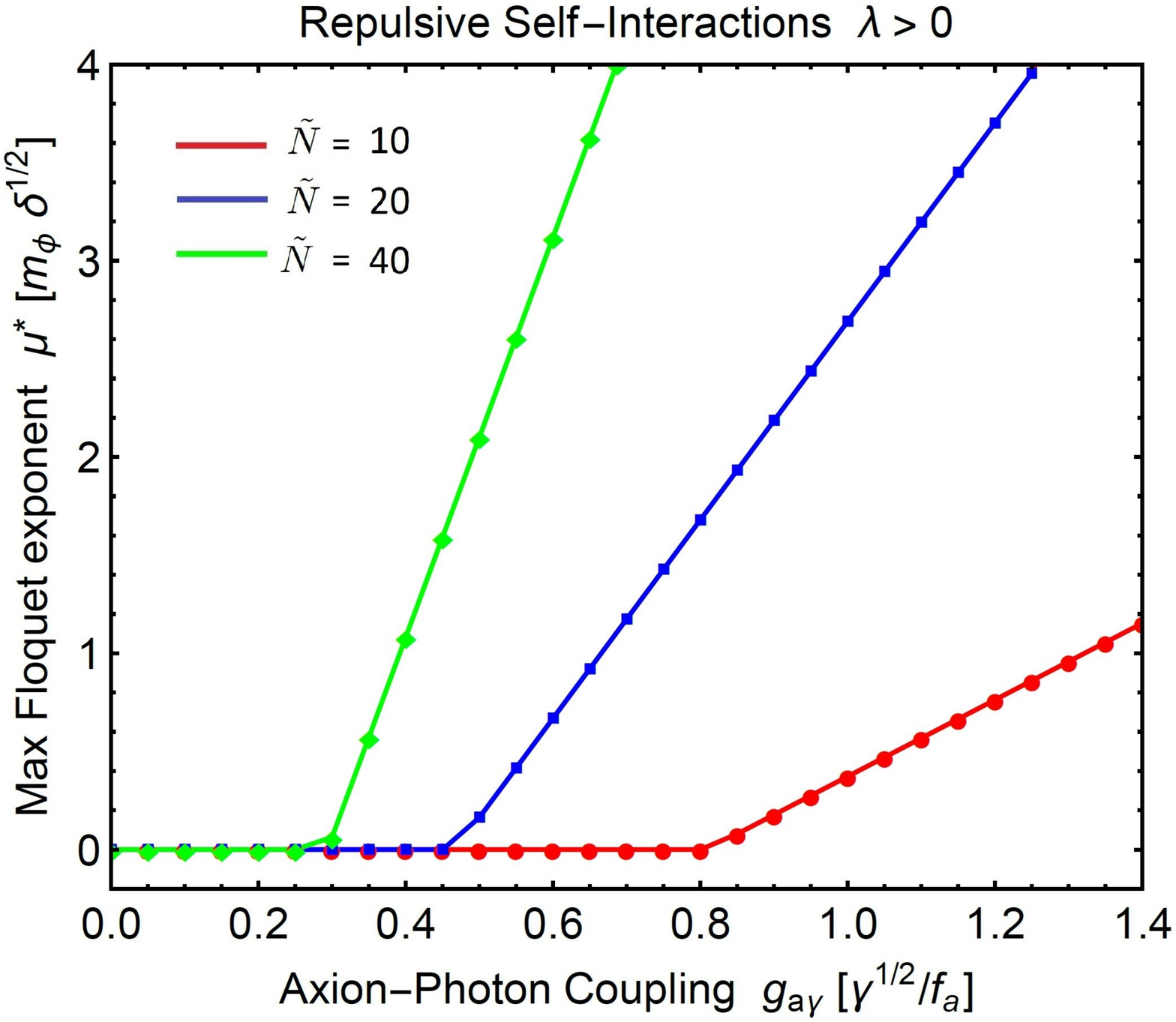}
\caption{The maximum real part of Floquet exponent $\mu_k^*$, describing parametric resonance of photons from a spherically symmetric clump condensate, as a function of axion-photon coupling $\ga$. We plot $\mu^*$ in units of $\ma\sqrt{\delta}$ and $\ga$ in units of $\sqrt{\gamma}/\fa$. Left panel: Attractive self-interactions $\lambda<0$ with $\tilde{N}=8$ in red, $\tilde{N}=9.2$ in blue, and $\tilde{N}=\tilde{N}_{max}\approx10.12$ in green, where $\tilde{N}=N/(|\lambda|\sqrt{\delta})$. Right panel: Repulsive self-interactions $\lambda>0$ with $\tilde{N}=10$ in red, $\tilde{N}=20$ in blue, and $\tilde{N}=40$ in green.}
\label{FigureFloquetCoupling}
\end{figure}

In Fig.~\ref{FigureFloquetCoupling} we show our results for the Floquet exponent as a function of axion-photon $\ga$ for 3 fixed values of the axion number $N$ for the clump solution. For the case of attractive self-interactions $\lambda<0$, recall there is a maximum axion number of $N_{max}\approx 10.12/(|\lambda|\sqrt{\delta})$. So we choose values of $N\leq N_{max}$. In the case of repulsive interactions we can choose larger values for $N$ as there is no maximum. We have measured the coupling $\ga$ in units of $\sqrt{|\lambda|}/\ma=\sqrt{\gamma}/\fa$, where $\gamma$ is the $\mathcal{O}(1)$ number that we mentioned in Eq.~(\ref{lambdachoice}) associated with the details of the axion potential; for the conventional QCD axion its preferred value is $\gamma\approx0.3$.

\begin{figure}[t]
\centering
\includegraphics[scale=0.3]{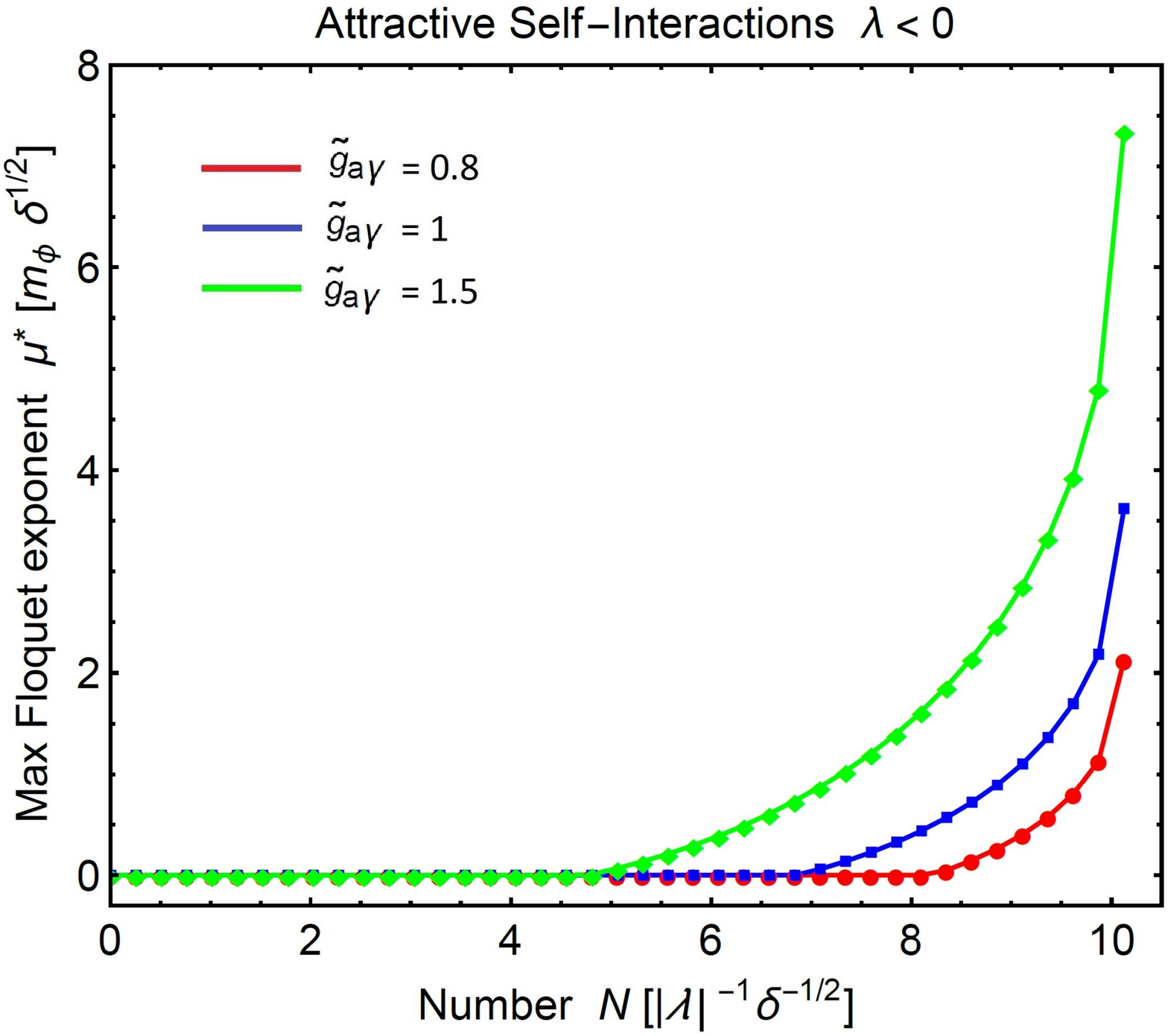}\!\!\!\!\!\!\!\!\!\!\!
\includegraphics[scale=0.3]{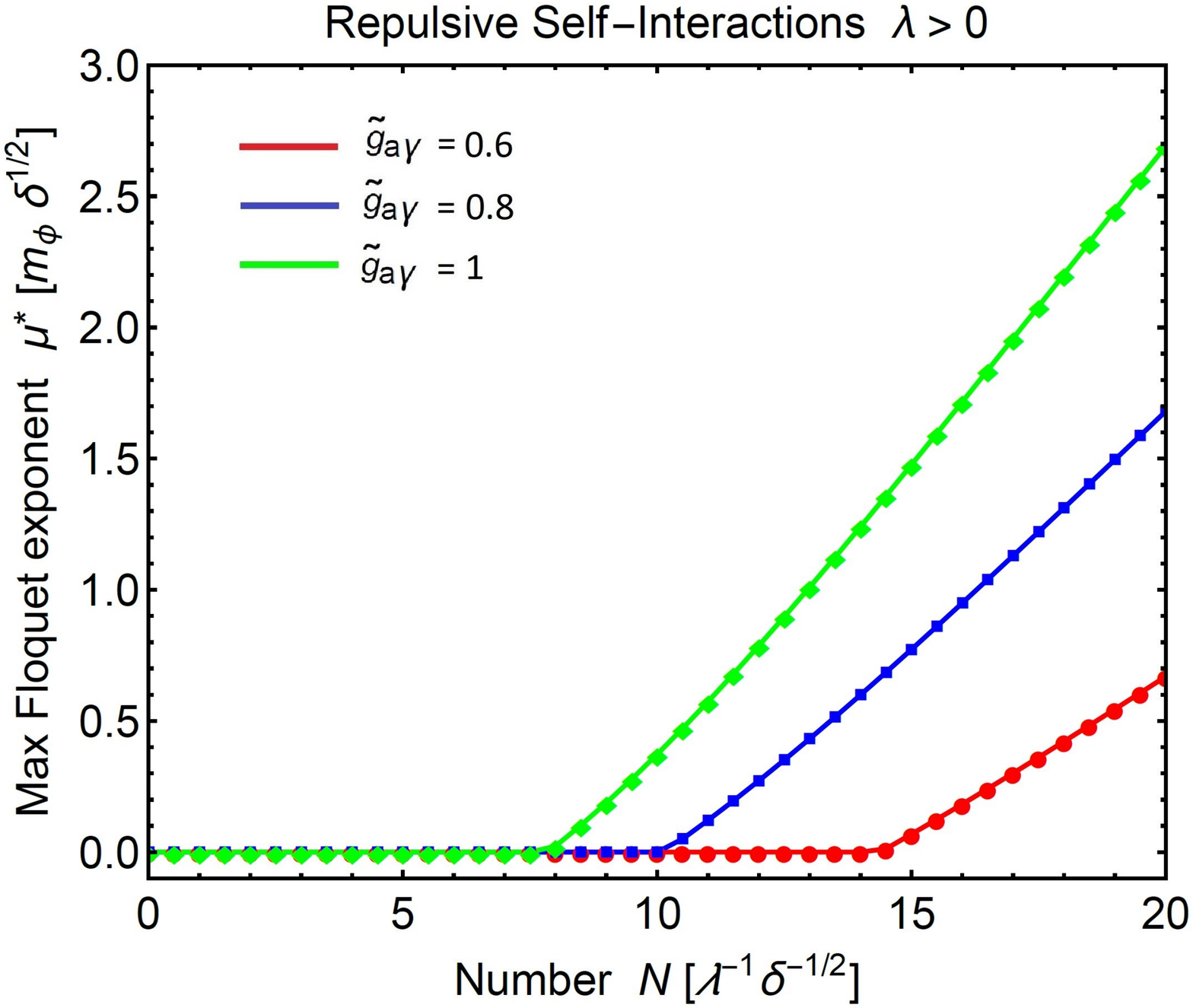}
\caption{The maximum real part of Floquet exponent $\mu_k^*$, describing parametric resonance of photons from a spherically symmetric clump condensate, as a function of clump number $N$. We plot $\mu^*$ in units of $\ma\sqrt{\delta}$ and $N$ in units of $|\lambda|^{-1/2}\delta^{-1/2}$. Left panel: Attractive self-interactions $\lambda<0$ with $\tga=0.8$ in red, $\tga=1$ in blue, and $\tga=1.5$ in green, where $\tga=\ga\,\fa/\sqrt{\gamma}$. Right panel: Repulsive self-interactions $\lambda>0$ with $\tga=0.6$ in red, $\tga=0.8$ in blue, and $\tga=1$ in green.}
\label{FigureFloquetNumber}
\end{figure}

In Fig.~\ref{FigureFloquetNumber} we show our results for a complementary analysis. Here we plot the Floquet exponent as a function of number $N$ for 3 fixed values of the axion-photon coupling $\ga$. 

For the QCD-axion with $\lambda<0$ and $\gamma=0.3$, we see from Fig.~\ref{FigureFloquetCoupling} left panel that when the axion clump has maximum particle number, we need the coupling to be greater than a minimum value
\beq
\ga > \gac = {\ba_c\over\fa}\,,\,\,\,\,\mbox{with}\,\,\,\,\ba_c\approx0.3\,,
\eeq
in order to have parametric resonance. In conventional QCD axion models, we expect $\ga=\mathcal{O}(10^{-2})/\fa$, so this would not be satisfied and there would be no resonance. On the other hand, for unconventional axion models, or for couplings to hidden sector photons (e.g., see Ref.~\cite{Daido:2018dmu}), this condition may be satisfied, leading to resonance. Alternatively, for repulsive self-interactions (e.g., see Ref.~\cite{Fan:2016rda}) there always exists sufficiently large $N$ to achieve resonance for any $\ga$ (unless $\ga$ is extremely small, in which case the required $N$ may be so large our non-relativistic approximations will eventually breakdown).

\section{General Criteria for Clump Resonance}\label{Criteria}

It is essential to notice a very big difference between the behavior of the Floquet exponent in the homogeneous condensate case of Section \ref{HomogeneousCondensates} and in the spatially localized clump case of Section \ref{SphericalCondensates}. In the homogenous case, there always exists a non-zero maximum Floquet exponent $\mu^*_H$, regardless of how small the axion-photon coupling $\ga$ is. Its value is proportional to $\ga$ (see Eq.~(\ref{FloquetMax}) and Fig.~\ref{FloquetHomogeneous}), and so long as $\ga$ is non-zero, then $\mu^*$ will be non-zero too, and there will be some resonance.

On the other hand, for a localized clump, as we saw in Figs.~\ref{FigureFloquetCoupling} \& \ref{FigureFloquetNumber}, the (real part of the) maximum Floquet exponent $\mu^*$ becomes strictly zero below a critical coupling $\ga$ or below a critical particle number $N$. 

There is in fact a very good physical reason for this. Imagine replacing the local clump condensate with central amplitude $\phiamp$ by the corresponding homogenous field configuration with the same amplitude $\phiamp$. If the width of the clump is very large, then the homogenous Floquet rate $\mu_H^*$ should approximate the clump Floquet rate $\mu^*$. On the other hand, if the width of the clump ($\approx2\,R$) is sufficiently small, then a produced pair of photons will escape the clump quicker than the time it would take for the next photon pair to be produced. In this case, Bose-Einstein statistics are ineffective as there are never more than $\mathcal{O}(1)$ occupancy number of photons within the clump to induce exponential growth, so the resonance will be shut-off leading to $\mu^*=0$ (it can still decay perturbatively into photons at the standard quantum decay rate of $\Gamma(\phi\to\gamma\gamma)=g_{a\gamma}^2\,\ma^3/(64\,\pi)$, which is much longer than the age of the universe for reasonable parameters of the QCD axion \cite{Raffelt:1985nk}). So the physical condition for parametric resonance to occur is that the homogenous growth rate $\mu^*_H\approx\ga\,\ma\,\phiamp/4$ is greater than the photon escape rate $\mu_{esc}\approx1/(2\,R)$
\beq
\mu_H^*>\mu_{esc}\,,\,\,\,\,(\mbox{resonance condition})\,.
\eeq
For earlier work on this idea see Ref.~\cite{Hertzberg:2010yz} where this criteria was originally discovered in the context of resonant scalar fields (also see Ref.~\cite{Kawasaki:2013awa}) and see Refs.~\cite{Tkachev:1986tr,Tkachev:1987cd,Tkachev:2014dpa} which discussed this in the context of the axion-photon resonance.

Furthermore, an excellent approximation to the growth rate from a localized clump is found to be the following
\beq
\mu^*\approx \bigg\{\!\begin{array}{l}\mu_H^* - \mu_{esc}\,,\,\,\,\,\mu_H^*>\mu_{esc}\,,\\ \,\,\,\,\,\,\,\,\,\,0\,,\,\,\,\,\,\,\,\,\,\,\,\,\,\,\,\,\mu_H^*<\mu_{esc}\,.\end{array}
\eeq
In the case of the above spherically symmetric clump configurations, the homogenous rate $\mu_H^*\approx\ga\,\ma\,\phiamp/4$ is evaluated by taking the amplitude as
\beq
\phiamp = \sqrt{2\over\ma}\,\Psi_0\,,\,\,\,\,\mbox{with}\,\,\,\Psi_0=\sqrt{3\,N\over\pi^3\,R^3}\,\,\,(\mbox{sech ansatz})\,.
\label{phiampsech}\eeq
We have confirmed the accuracy of this by computing both the exact numerical result for $\mu^*$ according to the prescription of Section \ref{SphericalCondensates} and this approximation for $\mu^*$, finding very close agreement. This approximate formula for $\mu^*$ is both very physical and very easy to evaluate, so we can use it in the next section to describe more complicated situations.

\section{Effective Photon Mass}\label{Photonmass}

In the previous sections we have considered massless photons. However, this picture is only rigorous in vacuum. In the not-quite-empty space of the interstellar medium, photons acquire an effective mass equal to the plasma frequency according to~\cite{Carlson:1994yqa}
\beq
\omega_p^2 = \frac{4 \pi \alpha\, n_e}{m_e} = \frac{n_e}{0.03\, \text{cm}^{-3}}\left( 6.4 \times 10^{-12}\, \text{eV} \right)^2\,,
\eeq
where $m_e$ and $n_e$ correspond to the mass and number density of the free electrons, respectively. In the very early universe, the number density of free electrons is so high that this photon plasma frequency is very high. This forbids resonance since the above decay processes $\phi\to \gamma+\gamma$ becomes kinematically forbidden. However, in the late universe, once clumps have formed, the number density of free electrons decreases. For the typical halo value today of $n_e\sim 0.03\,\text{cm}^{-3}$, we see that the plasma frequency is {\em much} smaller than the axion mass $m_a\sim 10^{-5}$\,eV, and so the process is easily kinematically allowed.

However, since we need to consider the fact that the electrons are spatially inhomogeneous, as they are moving in the galactic halo. To parameterize this we can write the plasma frequency as  $\omega_p(t) \approx \omega_p\, f(t)$, where $f(t)$ is a non-periodic time dependent function of $\mathcal{O}(1)$. To estimate the size of this effect, let us ignore the spatial structure for the moment, and focus on this new time dependence. The important fact is that if $f(t)$ is non-periodic then it could jeopardize the parametric resonance which relies on the existence of an (approximately) periodic pump. The modified equation of motion for the mode function $s_{k}$ is approximated as
\beq
{\ddot{s}_{\bf k}}+\left[k^2 + \omega_p^2(t)\,-\ga\,\om\,k\,\phiamp\sin(\om\,t)\right] {{s_{\bf{k}}}} = 0\,.
\eeq
In the small amplitude regime, the main contribution to the resonant process comes from the first instability band at $k \approx (\om/2)$. So, taking $\om \approx \ma$, $k \approx (\ma/2)$, $f_a \sim 6 \times 10^{11}\, \text{GeV}$, and $\ma \sim 10^{-5}\, \text{eV}$, we find that the ratio of the non-periodic plasma term to the periodic axion term is 
\beq
{\omega_p^2\over\ga\, \om\, k\, \phiamp} =\mathcal{O}(10^{-4})\,.
\eeq
Hence the plasma mass corrections are expected to be negligible, indicating that our above massless photon approximation is reasonable. Further analysis of this topic may be useful.

\section{Clump Condensates with Angular Momentum}\label{AngularCondensates}

In Ref.~\cite{Hertzberg:2018lmt} we studied BECs with non-zero angular momentum. We found that such clumps have a larger maximum particle number (for attractive self-interactions) and therefore these types of clumps may be better for achieving resonance as the field amplitude will also be larger. We investigate this possibility here.

\subsection{Non-Spherical Clump Profile}

As we did in Ref.~\cite{Hertzberg:2018lmt}, we take the field profile to factorize into a radial profile $\Psi(r)$ and a single spherical harmonic $Y_{lm}$ as follows
\beq
\psi({\bf x},t) =\sqrt{4\pi}\,\Psi(r)\,Y_{lm}(\theta,\varphi)\,e^{-i\,\mu\,t}\,,
\label{trialangular}
\eeq
(note that the italic indices $\{l,m\}$ here refer to the axion field, and not to be confused with the non-italic indices $\{\La,\Ma\}$ that referred to the electromagnetic field in Section \ref{Vector}). This ansatz is not exact since non-linearities will couple different spherical harmonics, but it will suffice for our purposes here. The corresponding angular momentum is
\beq
{\bf L}=(0,0,N m)\,,
\label{AngMom}
\eeq
(with $N=4\pi\int_0^\infty dr\, r^2|\Psi(r)^2|$) and evidently only depends on the spherical harmonic number $m$ and not $l$. As we showed in Ref.~\cite{Hertzberg:2018lmt}, states that minimize the energy at fixed particle number $N$ and fixed angular momentum $L_z=N\,m$ occur for $l=|m|$, which we shall focus on.

Following Ref.~\cite{Hertzberg:2018lmt} we note that the above sech profile that we used for the spherically symmetric case is not as accurate for states with non-zero angular momentum. In particular, the equation of motion demands that the field around $r=0$ has the form $\Psi(r)=\Psi_{\alpha}\,r^l-{1\over 2}\Psi_{\beta}\,r^{l+2}+\ldots$. A very useful approximate form of the solution that obeys this property and is known to be quite accurate numerically, is to take the profile to a modified Gaussian
\beq
\Psi(r)=\sqrt{N\over2\pi(l+{1\over2})! \, R^3}\,\left(r\over R\right)^{\!l}\,e^{-r^2/(2 R^2)}\,\,\,\,\,(\mbox{modified Gaussian ansatz})\,,
\eeq
where the (variational) radius $R$ can now be a function of $\{l,m\}$. 

By inserting this into the Hamiltonian Eq.~(\ref{HamTotal}) and integrating we find a generalization of Eq.~(\ref{HamSech}) which allows for non-zero angular momentum
\beq
H(R) = a_l\frac{N}{2\ma\,R^2} - b_{lm}\frac{G\,\ma^2\,N^2}{R} +c_{lm}\frac{\lambda\,N^2}{\ma^2\,R^3 }\,.
\label{HamAng}
\eeq
In this modified Gaussian ansatz the form of $a_l$ is simple, but the exact forms of $b_{lm}$ and $c_{lm}$ are complicated and are reported in the Appendix. It suffices to report on their approximate values for high angular momentum $|m|=l$
\beq
a_l\approx{l\over2}\,,\,\,\,\,b_{lm}  \approx  0.336\sqrt{\ln l\over l}\,,\,\,\,\,c_{lm}  \approx {1\over 32\pi^2\sqrt{2\,l}}\,,\,\,\,\,(\mbox{high}\,\,|m|=l)\,.
\eeq

For attractive self-interactions $\lambda<0$ there is once again a maximum number of allowed particles in a clump, which is a generalization of the spherically symmetric result in Eq.~(\ref{Nmax}) to
\beq
N_{max} = {a_l\over\sqrt{3\,b_{lm}\,c_{lm}}}{1\over|\lambda|\sqrt{\delta}} \approx {10.52\,l^{3/2}\over(\ln l)^{1/4}}\,,\,\,\,\,(\mbox{high}\,\,|m|=l)\,.
\label{NmaxAng}\eeq
Since $N_{max}$ is rapidly increasing with $l$ the field amplitude is quite large and therefore there is an increased chance of resonance into photons which we explore in the next subsection.

\subsection{Approximate Treatment of Clump Resonance}

For a non-spherically symmetric clump configuration in principle we should return to the full equation of motion for the photon (\ref{EMeom}) in order to compute its behavior. However, the problem is now fully 3-dimensional and we cannot reduce it to an effective 1-dimensional problem as we did in Section \ref{Dominant} when we studied spherically symmetric clumps. Although such a problem may still be doable numerically, it suffices to utilize the important result established in Section \ref{Criteria}, where it was explained that the condition for resonance is that the maximum homogenous Floquet exponent $\mu_H^*$ is larger than the effective escape rate $\mu_{esc}$. We expect this basic idea carries over to non-spherical pump configurations, and so we shall use that idea in this section.

Firstly, we need to determine the maximum homogeneous Floquet exponent $\mu_H^*$. Recall that it is proportional to the field's amplitude $\phiamp$ as $\mu_H^*\approx\ga\,\ma\,\phiamp/4$. The field amplitude is given by the following generalization of Eq.~(\ref{phiampsech}) to include a spherical harmonic
\beq
\phiamp = \sqrt{2\over\ma}\,\Psi_0\sqrt{4\,\pi}\,|Y_{lm}|_0\,.
\eeq
Here the maximum value or amplitude $\Psi_0$ of the modified Gaussian occurs at a radius of $r=\sqrt{l}\,R$ with value
\beq
\Psi_0 = \sqrt{N\over R^3}\,f_l,\,\,\,\,\,\mbox{with}\,\,\,\,
f_l = \sqrt{l^{l}\,e^{-l}\over2\pi(l+{1\over2})!} \approx {1\over(2\pi)^{3/4}\,\sqrt{l}}\,,\,\,\,\,(\mbox{high}\,\,l)\,.
\eeq
Also the maximum value $|Y_{lm}|_0$ of the spherical harmonic with $l=|m|$ occurs at a polar angle of $\theta=\pi/2$ with value
\beq
\sqrt{4\pi}\,|Y_{lm}|_0={\sqrt{(2l+1)!}\over 2^ll!}\approx {\sqrt{2}\,l^{1/4}\over\pi^{1/4}}\,,\,\,\,\,(\mbox{high}\,\,|m|=l)\,.
\eeq

For attractive self-interactions (which is expected for axions) the best possibility for resonance is when $N$ is taken to its maximum value $N_{max}$ given above in Eq.~(\ref{NmaxAng}), with a corresponding (minimum) radius $R_{min}=\sqrt{3\,c_{lm}/b_{lm}}/(\ma\sqrt{\delta})$. By taking $N\to N_{max}$ and evaluating the above expression for $\Psi_0$ we find the following form for the maximum Floquet exponent of the corresponding homogeneous condensate
\beq
\mu_H^*\approx 5.8\,\sqrt{l}\,(\ln l)^{1/4}\,\tga\,\ma\sqrt{\delta}\,,\,\,\,\,(N=N_{max}\,;\,\,\,\mbox{high}\,\,|m|=l)\,,
\label{muHhighl}\eeq
where $\tga\equiv\ga\,\fa/\sqrt{\gamma}$.

Secondly, we need to determine the effective escape rate of the photons $\mu_{esc}$. As we explained in Section \ref{Criteria} this is set by the inverse width of the axion clump. For a simple spherically symmetric clump, like the sech function we studied earlier, this is clearly $\mu_{esc}\approx 1/(2\,R)$, where $R$ is the argument of the sech function. However for these non-spherical clumps there are potentially several complications. In particular, the radial profile is not peaked around $r=0$ for non-zero $l$. Instead it is peaked around $R_p=\sqrt{l}\,R$. In fact the modified Gaussian becomes an ordinary Gaussian with a shifted peak at large $l$ as
\beq
\Psi(r)\approx \sqrt{N\over(2\pi)^{3/2}\,l\,R^3}\,e^{-(r-R_p)^2/R^2}\,,\,\,\,\,(\mbox{high}\,\,l)\,,
\eeq
Notice that its full width in this radial direction is still of the order $w_{r}\sim 2\,R$. 

Also, the angular dependence provided by the spherical harmonic is somewhat non-trivial. For $l=|m|$ the spherical harmonic is $Y_{lm}(\theta,\varphi)=e^{\pm i\,l\,\varphi}(-\sin\theta)^l|Y_{lm}|_0$. Then when we form the real field $\phi$ we obtain
\beq
\phi(r,\theta,\varphi,t)=\sqrt{2\over\ma}\,\sqrt{4\,\pi}|Y_{lm}|_0\, \Psi(r)\,(-\sin\theta)^l\cos(\om\,t\pm l\,\varphi)\,,
\eeq 
We see that the azimuthal angle $\varphi$ acts as a phase-shift of the periodic oscillations in time. This does not appear to appreciably alter the coherence of the pump field, and so we anticipate that it does not affect the resonant structure appreciably. On the other hand, the dependence on the polar variable $\theta$ implies that the field is peaked at $\theta=\pi/2$ with a full width half maximum of 
\beq
\Delta\theta = 2\cos^{-1}\!\left(e^{-\frac{1}{l}\ln 2}\right)\,.
\label{deltathetaii}
\eeq
For $l\to0$ this recovers $\Delta\theta=\pi$, while for large $l$ it decreases as $\Delta\theta\approx2\sqrt{2\ln2/l}$. This implies that the full width in this polar direction is $w_\theta\sim\Delta\theta\,R_p/\pi\sim 2\,R$ (using $R_p=\sqrt{l}\,R$), which is of the same order as $w_r$. 

Hence, although it is not exact, we can estimate the effective width of the clump as simply $\sim 2\,R$ and so we take the escape rate as $\mu_{esc}\approx1/(2\,R)$ as we did in the spherically symmetric case. For attractive self-interactions, we again take $N\to N_{max}$ and we obtain
\beq
\mu_{esc}\approx 3.5\,(\ln\,l)^{1/4}\,\ma\sqrt{\delta} \,,\,\,\,\,(N=N_{max}\,;\,\,\,\mbox{high}\,\,|m|=l)\,.
\label{mueschighl}\eeq

\begin{figure}[t]
\centering
\includegraphics[scale=0.34]{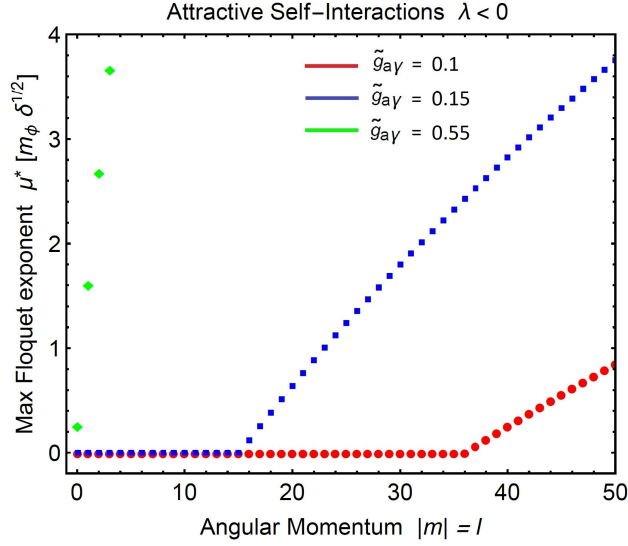}
\caption{The maximum real part of Floquet exponent $\mu_k^*$, describing parametric resonance of photons from a clump condensate as a function of its angular momentum $|m|=l$. We plot $\mu^*$ in units of $\ma\sqrt{\delta}$. This is for attractive self-interactions with $N=N_{max}$. Here $\tga=0.1$ in red, $\tga=0.15$ in blue, and $\tga=0.55$ in green, where $\tga=\ga\,\fa/\sqrt{\gamma}$.}
\label{FigureFloquetAngular}
\end{figure}

Using the approximate formula for the maximum Floquet exponent $\mu^*$ from Section \ref{Criteria} as $\mu^*\approx\mbox{Max}\{\mu_H^*-\mu_{esc},0\}$ we have evaluated this for $l=0,1,\ldots,50$ and plotted the result in Fig.~\ref{FigureFloquetAngular} (we have computed these quantities exactly rather than just using the high $|m|=l$ approximations).

Since $\mu_H^*$ in Eq.~(\ref{muHhighl}) grows with angular momentum $|m|=l$ faster than $\mu_{esc}$ in Eq.~(\ref{mueschighl}) does, it becomes easier to achieve resonance for BECs of higher angular momentum. 

\begin{figure}[t]
\centering
\includegraphics[scale=0.39]{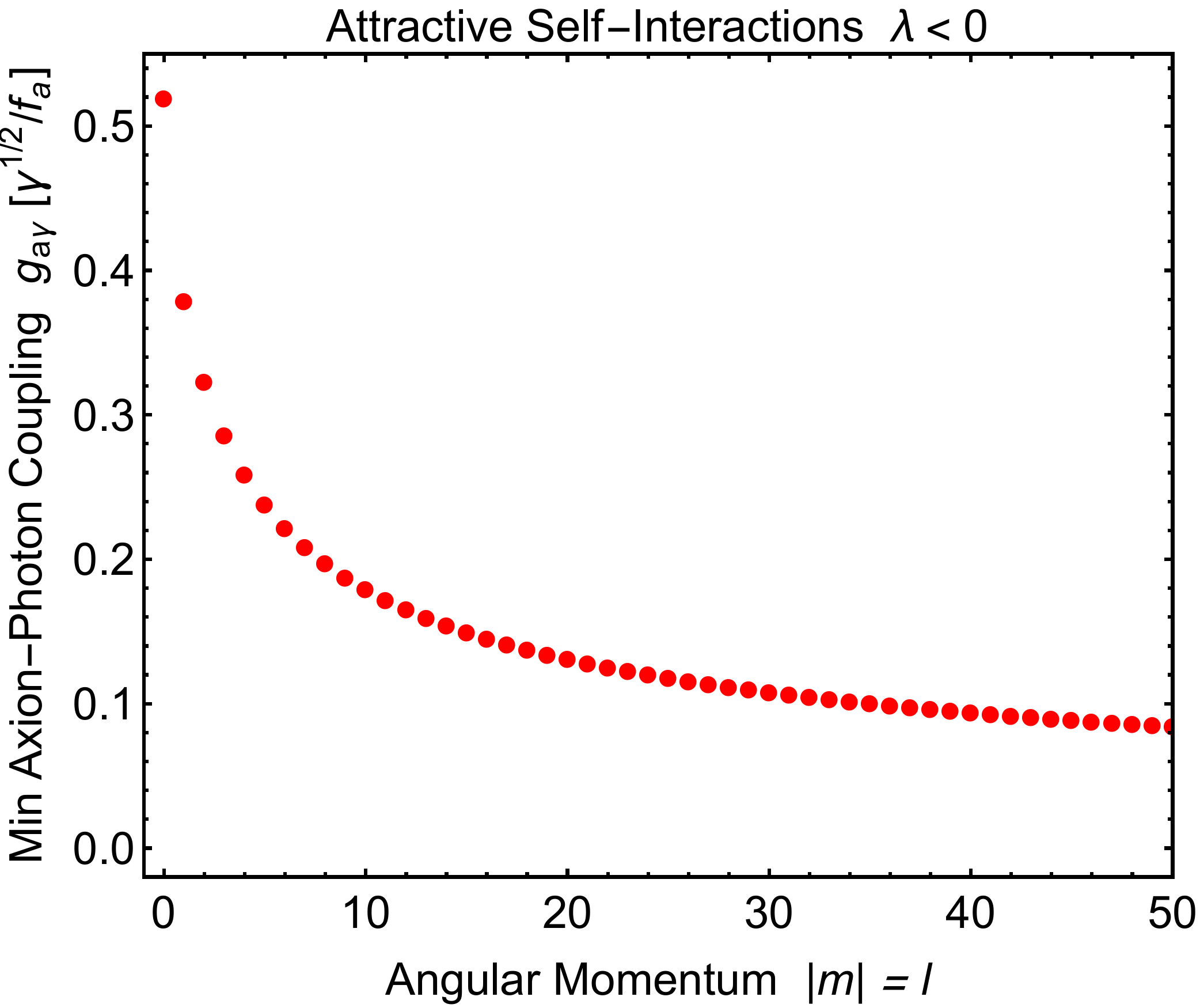}
\caption{The minimum axion-photon coupling $\ga$ that is necessary in order to have resonance from a clump condensate as a function of its angular momentum $|m|=l$. We plot $\ga$ in units of $\sqrt{\gamma}/\fa$. This is for attractive self-interactions with $N=N_{max}$.}
\label{FigureCouplingAngular}
\end{figure}

In fact by equating $\mu_H^*=\mu_{esc}$, we can determine the minimum axion-photon coupling $\ga$ that will allow at least some clumps (namely those with $N\approx N_{max}$) to undergo parametric resonance into photons. We plot this in Fig.~\ref{FigureCouplingAngular}. In both Figs.~\ref{FigureFloquetAngular} \& \ref{FigureCouplingAngular} the $l=0$ value is obtained from our previous spherically symmetric analysis using the sech profile, while for $l\ge1$ we use the modified Gaussian profile discussed here.
For high angular momentum, the minimum axion-photon coupling that can achieve resonance is easily found from Eqs.~(\ref{muHhighl},\,\ref{mueschighl}). By taking $\gamma=0.3$ we obtain
\beq
\ga > \gac =  {\ba_c\over\fa}\,,\,\,\,\,\mbox{with}\,\,\,\,\ba_c\approx{0.3\over\sqrt{l}}\,,\,\,\,\,(N=N_{max}\,;\,\,\,\mbox{high}\,\,|m|=l)\,.
\eeq
Hence in order to achieve resonance with conventional values of $\ga$ for the QCD axion $\ga=\mathcal{O}(10^{-2})/\fa$, we need rather large angular momentum of $|m|=l\gtrsim\mathcal{O}(10^3)$.

\section{Astrophysical Consequences and Discussion}\label{Astrophysical}

We have seen that under certain conditions, parametric resonance of axion clump condensates into photons is possible. Suppose that the axion-photon coupling $\ga$ is greater than the minimum value discussed above. This will lead to an exponential growth in the electromagnetic field. (Even though quantum fluctuations in the initial electromagnetic field should ensure that the field has an initial non-zero value, we can also consider that axion clumps in the galactic halo are immersed in a bath of several electromagnetic radiations coming from astrophysical sources and the CMB. In any case, there should be no inconvenience so that both quantum effects and real photons act as seeds for exponential growth over time.) The photon occupancy number then increases from small to much larger values, which implies that the final output is an essentially classical electromagnetic wave. We can estimate the time-scale for this growth for typical values of the QCD axion. When $\mu_H^*$ is somewhat larger than $\mu_{esc}$, we can just use the homogenous formula estimate $\mu^*\sim\mu_H^*\approx\ga\,\ma\,\phiamp/4$. Let us consider the true BEC ground states, which are spherically symmetric. For $N\sim N_{max}$ (attractive self-interactions) this rate is on the order 
\beq
\mu^*\sim 15\,\ga\,\fa\,\ma\,\sqrt{\delta}\gtrsim 5\,\ma\sqrt{\delta}\,,
\eeq 
where in the last step we have used our earlier result $\ga\,\fa>0.3$ for resonance. For the QCD axion the expected value of $\delta=G\,\fa^2/\gamma$ is very small. As an example, for $\fa\sim 6\times10^{11}$\,GeV, we have $\delta\sim 5\times10^{-15}$. For an axion mass of $\ma\sim 10^{-5}$\,eV, this gives a growth time-scale of $\tau=1/\mu^*\lesssim 2\times 10^{-4}$\,sec. This is a very short time-scale for an astrophysical process. There are two possible consequences of this, which we now discuss.

\subsection{Clump Mass Pile-Up}

\begin{figure}[t]
\centering
\includegraphics[scale=0.4]{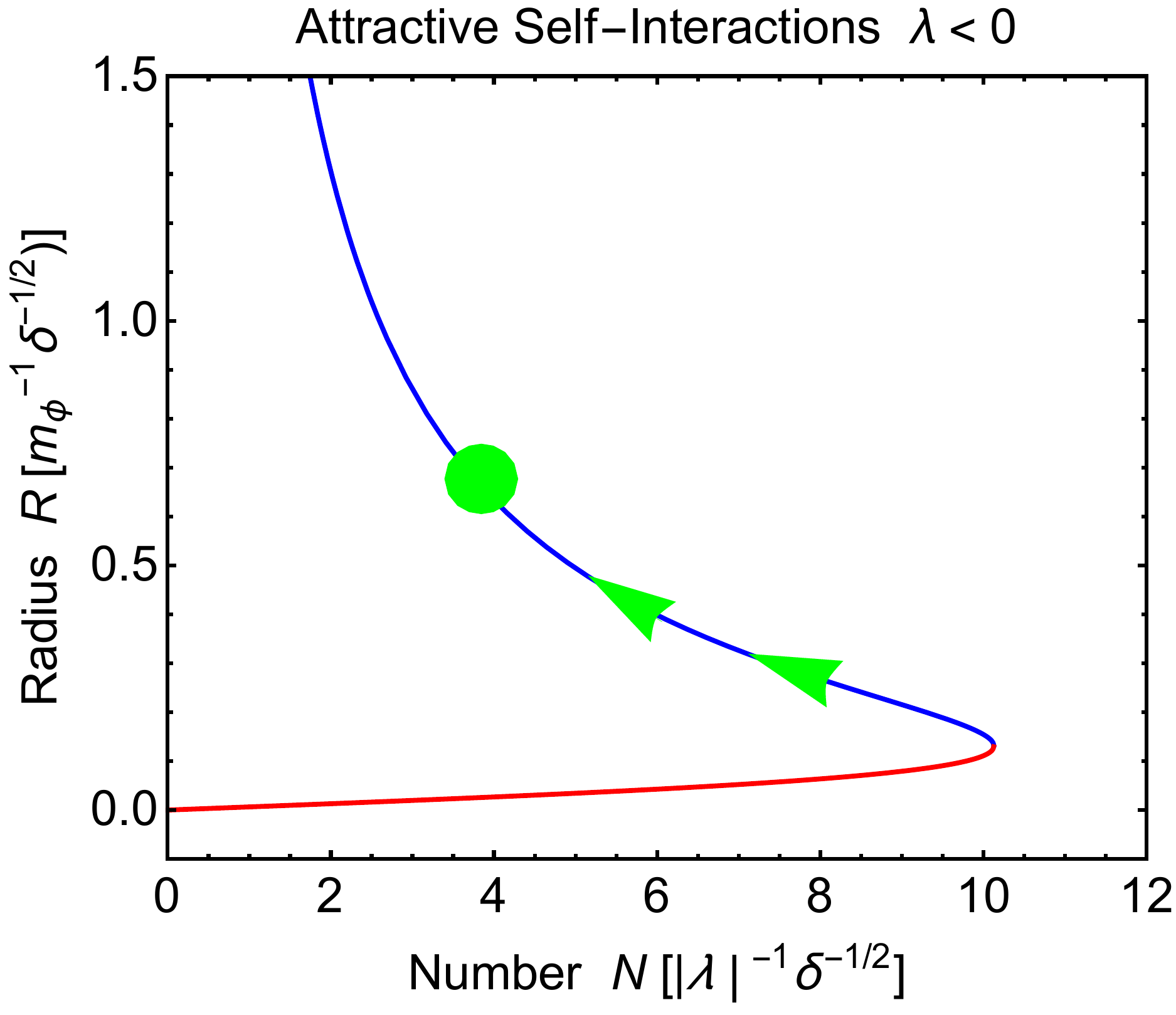}
\caption{Clump radius $R$ as a function of clump number $N$ for spherically symmetric clumps with attractive self-interactions $\lambda<0$. We have taken the axion-photon coupling to be $\tga=\ga\,\fa/\sqrt{\gamma}=2$ here. For any clumps on the stable blue branch with number $N>N_c$ they will resonantly produce photons, lose mass, and pile-up at the critical value $N_c\approx3.7/(|\lambda|\sqrt{\delta})$.}
\label{FigureRadiusNumberPileUp}
\end{figure}

After the QCD phase transition in the early universe, the axion field begins red-shifting. After some time this means its field amplitude is sufficiently small that it appears to not satisfy the conditions for resonance. However, in the later universe, once gravitational interactions become appreciable the axion field can potentially undergo gravitational thermalization forming these clumps throughout the universe. Suppose the axion-photon coupling $\ga$ is appreciable. Then as the axion clumps are forming, some of them will achieve a sufficiently large mass and hence field amplitude that they can undergo parametric resonance into photons. Such clumps would then quickly lost energy into electromagnetic radiation, causing the clump's mass to decrease. This would continue until the clump's field amplitude is sufficiently small that the resonance is shut-off. As we showed earlier, there exists a critical value for resonance. Fig.~\ref{FigureFloquetNumber} shows that for a fixed value of the axion-photon coupling, there is a critical number $N_c$ that allows for resonance. So if the clump begins with number $N>N_c$ it will radiate into photons, losing mass until $M\to N_c\,\ma$ and the resonance will stop. We illustrate this idea in Fig.~\ref{FigureRadiusNumberPileUp} where we describe spherically symmetric condensate clumps with $\tga=\ga\,\fa/\sqrt{\gamma}=2$, which corresponds to a critical number $N_c\approx 3.7\,/(|\lambda|\sqrt{\delta})$. We have used green arrows to indicate that for initial numbers larger than this, they will flow towards lower values, leading to a pile-up at a unique value indicated by the green circle. It is interesting to note that this critical mass $M_c=N_c\,\ma$ is determined purely in terms of fundamental constants. If such a pile-up of masses were detected it would be a clear signature of the axion model.

\subsection{Electromagnetic Emission in the Sky}

The above idea is focussed on resonant processes that would presumably have occurred in the distant past when the axion clumps first form. However we can imagine a scenario where the process is still occurring. Consider a pair of clump condensates, each with number $N_1$ and $N_2$ with $N_1<N_c$ and $N_2<N_c$, where $N_c$ is the minimum number required for resonance. Now suppose that these clumps happen to merge together in the late universe. If the total number $N_{tot}=N_1+N_2>N_c$ (which is plausible given the pile-up near $N_c$ mentioned above) then resonance will suddenly begin to occur, driving the total towards $N_{tot}\to N_c$. This would lead to a sudden significant emission of electromagnetic radiation in the galaxy. The corresponding electromagnetic output would be a narrow line near the resonant wavelength of $\lambda_{EM}\approx 2\pi/\om\approx4\pi/\ma$. For an axion mass of $\ma\sim 10^{-5}$\,eV, this is a radio wave line $\lambda_{EM}\sim$\,$10^{-1}$m.

Now the typical mass of a clump is on the order $\sim 10^{-11}\,M_\odot$, which is comparable to the moon's mass. If the merger took place, an amount of energy comparable to $M\,c^2$ equivalent of the moon's mass would be emitted into the galaxy. While this may be a small amount of energy compared to, say, a typical supernovae explosion and may be a slower process, it is possible that it would occur more frequently if these axion clumps comprise a significant fraction of the dark matter of the universe. However we leave an estimate of the merger rate and corresponding detection viability for future work.

\subsection{Outlook}

In this work we have explored a possible novel consequence of the axion model, in which gravitationally bound axion clumps can form and, under certain conditions, can undergo parametric resonance into electromagnetic radiation. For conventional values of axion-photon coupling this is impossible to achieve for an ordinary axion with attractive self-interactions in its BEC ground state. However it can occur for a condensate of sufficiently large angular momentum. Furthermore, it can occur for atypically large axion-photon coupling $\ga\gtrsim1/\fa$, as well as for couplings to hidden sector photons in which the coupling is essentially unconstrained, as well as for scalars with repulsive self-interactions since there is no maximum clump mass in this case. 

It would be interesting to further explore possible theoretical realizations of these more general possibilities (e.g., see Refs.~\cite{Daido:2018dmu,Fan:2016rda}) and to compare this to existing bounds on the axion. It would also be worthwhile to explore possible hints of the above ideas of a clump mass pile-up and of these sudden electromagnetic emissions. It is especially important to numerically compute the clump abundance and merger rate. Such possible astrophysical consequences of axions and axion-like-particles is an interesting direction that may help to unravel the nature of dark matter.

\section*{Acknowledgments}

MPH is supported by National Science Foundation grant PHY-1720332.

\appendix

\section{Hamiltonian with Angular Momentum}\label{Appendix}

The effective Hamiltonian for states of non-zero angular momentum Eq.~(\ref{HamAng}) is specified by the following coefficients in the modified Gaussian ansatz
\bea
a_{l}\amp = \amp {3+2l\over4}\,, \\
b_{lm}\amp = \amp \sum_{l'=0}^{2l}\Clm(l')\,J_l(l')\,, \\
c_{lm}\amp = \amp {(2l+{1\over2})!\over 2^{2l+{13\over2}}\pi\,[(l+{1\over2})!]^2}\sum_{l'=0}^{2l}(2l'+1)\Clm(l')\,.
\eea
Here the coefficients $\Clm(l')$ are the Wigner 3-j symbols 
\beq
\Clm(l') = (2l+1)^2\binom {l~~l'~~l} {0~~0~~0}^{\!2} \binom {~l~~l'~~l} {\!-m~0~~m}^{\!2}\,,
\eeq
and the coefficients $J_l(l')$ are related to hypergeometric functions as follows
\bea
J_l(l') \amp = \amp {(2l+{3\over2})!\, _2F_1(2l+{5\over2},l+{l'+3\over2},l+{l'+5\over2},-1)\over(3+2l+l')[(l+{1\over2})!]^2}\nonumber\\
\amp+\amp{(l-{l'\over2})!\,[(l+{l'+1\over2})!-(2l+{3\over2})!\, _2F_1^{reg}(2l+{5\over2},l-{l'\over2}+1,l-{l'\over2}+2,-1)]\over2[(l+{1\over2})!]^2}\,.
\eea


\begin{thebibliography}{00}

\bibitem{Schiappacasse:2017ham} 
  E.~D.~Schiappacasse and M.~P.~Hertzberg,
  ``Analysis of Dark Matter Axion Clumps with Spherical Symmetry,''
  JCAP {\bf 1801}, 037 (2018)
  Erratum: [JCAP {\bf 1803}, no. 03, E01 (2018)]
  [arXiv:1710.04729 [hep-ph]].
%
\bibitem{Hertzberg:2018lmt} 
  M.~P.~Hertzberg and E.~D.~Schiappacasse,
  ``Scalar Dark Matter Clumps with Angular Momentum,''
  arXiv:1804.07255 [hep-ph].

\bibitem{Peebles:2013hla} 
  P.~J.~E.~Peebles,
  ``Dark Matter,''
  arXiv:1305.6859 [astro-ph.CO].
 
\bibitem{Preskill:1982cy} 
  J.~Preskill, M.~B.~Wise and F.~Wilczek,
  ``Cosmology of the Invisible Axion,''
  Phys.\ Lett.\ B {\bf 120}, 127 (1983).
%
\bibitem{Abbott:1982af} 
  L.~F.~Abbott and P.~Sikivie,
  ``A Cosmological Bound on the Invisible Axion,''
  Phys.\ Lett.\ B {\bf 120}, 133 (1983).
%
\bibitem{Dine:1982ah} 
  M.~Dine and W.~Fischler,
  ``The Not So Harmless Axion,''
  Phys.\ Lett.\ B {\bf 120}, 137 (1983).
%
\bibitem{Kim:2008hd} 
  J.~E.~Kim and G.~Carosi,
  ``Axions and the Strong CP Problem,''
  Rev.\ Mod.\ Phys.\  {\bf 82}, 557 (2010)
  [arXiv:0807.3125 [hep-ph]].

\bibitem{Peccei:1977hh} 
  R.~D.~Peccei and H.~R.~Quinn,
  ``CP Conservation in the Presence of Instantons,''
  Phys.\ Rev.\ Lett.\  {\bf 38}, 1440 (1977).
%
\bibitem{Weinberg:1977ma} 
  S.~Weinberg,
  ``A New Light Boson?,''
  Phys.\ Rev.\ Lett.\  {\bf 40}, 223 (1978).
%
\bibitem{Wilczek:1977pj} 
  F.~Wilczek,
  ``Problem of Strong P and T Invariance in the Presence of Instantons,''
  Phys.\ Rev.\ Lett.\  {\bf 40}, 279 (1978).

\bibitem{Hertzberg:2008wr} 
  M.~P.~Hertzberg, M.~Tegmark and F.~Wilczek,
  ``Axion Cosmology and the Energy Scale of Inflation,''
  Phys.\ Rev.\ D {\bf 78}, 083507 (2008)
  [arXiv:0807.1726 [astro-ph]].

\bibitem{Asztalos:2009yp} 
  S.~J.~Asztalos {\it et al.}  [ADMX Collaboration],
  ``A SQUID-based microwave cavity search for dark-matter axions,''
  Phys.\ Rev.\ Lett.\  {\bf 104}, 041301 (2010)
  [arXiv:0910.5914 [astro-ph.CO]].
%
\bibitem{Hoskins:2011iv} 
  J.~Hoskins, J.~Hwang, C.~Martin, P.~Sikivie, N.~S.~Sullivan, D.~B.~Tanner, M.~Hotz and L.~J.~Rosenberg {\it et al.},
  ``A search for non-virialized axionic dark matter,''
  Phys.\ Rev.\ D {\bf 84}, 121302 (2011)
  [arXiv:1109.4128 [astro-ph.CO]].

\bibitem{Sikivie:2009qn} 
  P.~Sikivie and Q.~Yang,
  ``Bose-Einstein Condensation of Dark Matter Axions,''
  Phys.\ Rev.\ Lett.\  {\bf 103}, 111301 (2009)
  [arXiv:0901.1106 [hep-ph]].
%
\bibitem{Erken:2011dz} 
  O.~Erken, P.~Sikivie, H.~Tam and Q.~Yang,
  ``Cosmic axion thermalization,''
  Phys.\ Rev.\ D {\bf 85}, 063520 (2012)
  [arXiv:1111.1157 [astro-ph.CO]].

\bibitem{Guth:2014hsa} 
  A.~H.~Guth, M.~P.~Hertzberg and C.~Prescod-Weinstein,
  ``Do Dark Matter Axions Form a Condensate with Long-Range Correlation?,''
  Phys.\ Rev.\ D {\bf 92}, no. 10, 103513 (2015)
  [arXiv:1412.5930 [astro-ph.CO]].

\bibitem{Kofman:1994rk}
L.~Kofman, A.~Linde and A.~A.~Starobinsky,
 ``Reheating after Inflation,''
Phys.\ Rev.\ Lett.\ {\bf 73}, 3195 (1994)
[arXiv:hep-th/9405187].
%
\bibitem{Shtanov:1995}
Y.~Shtanov, J.~Traschen R.~Brandenberger
 ``Universe reheating after inflation,''
Phys.\ Rev.\ D\ {\bf 51}, 5438 (1995)
[arXiv:hep-ph/9407247].

\bibitem{Espriu:2011vj}
D.~Espriu and A.~Renau,
 ``Photon propagation in a cold axion background with and without magnetic field,''
Phys.\ Rev.\ D.\ {\bf {85}}, 025010 (2012)
[arXiv:1106.1662 [hep-ph]].

\bibitem{Tkachev:1986tr} 
  I.~I.~Tkachev,
  ``Coherent scalar field oscillations forming compact astrophysical objects,''
  Sov.\ Astron.\ Lett.\  {\bf 12}, 305 (1986)
  [Pisma Astron.\ Zh.\  {\bf 12}, 726 (1986)].
%
\bibitem{Tkachev:1987cd} 
  I.~I.~Tkachev,
  ``An Axionic Laser in the Center of a Galaxy?,''
  Phys.\ Lett.\ B {\bf 191}, 41 (1987).
%
\bibitem{Tkachev:2014dpa} 
  I.~I.~Tkachev,
  ``Fast Radio Bursts and Axion Miniclusters,''
  JETP Lett.\  {\bf 101}, no. 1, 1 (2015)
  [Pisma Zh.\ Eksp.\ Teor.\ Fiz.\  {\bf 101}, no. 1, 3 (2015)]
  [arXiv:1411.3900 [astro-ph.HE]].

 \bibitem{Yoshimura:1995gc}
  M.~Yoshimura,
  ``Catastrophic particle production under periodic perturbation,''
  Prog.\ Theor.\ Phys.\  {\bf 94} (1995) 873
  [arXiv:hep-th/9506176].
  %
  \bibitem{Yoshimura:1996fk}
  M.~Yoshimura,
  ``Decay rate of coherent field oscillation,''
  arXiv:hep-ph/9603356 (1996).
  %
  \bibitem{Kitajima:2017peg} 
  N.~Kitajima, T.~Sekiguchi and F.~Takahashi,
  ``Cosmological abundance of the QCD axion coupled to hidden photons,''
  arXiv:1711.06590 [hep-ph].

\bibitem{Hertzberg:2016tal} 
  M.~P.~Hertzberg,
  ``Quantum and Classical Behavior in Interacting Bosonic Systems,''
  JCAP {\bf 1611}, no. 11, 037 (2016)
  [arXiv:1609.01342 [hep-ph]].  
  
\bibitem{diCortona:2015ldu} 
  G.~Grilli di Cortona, E.~Hardy, J.~Pardo Vega and G.~Villadoro,
  ``The QCD axion, precisely,''
  JHEP {\bf 1601}, 034 (2016)
  [arXiv:1511.02867 [hep-ph]].
  
\bibitem{Kolb:1993zz} 
  E.~W.~Kolb and I.~I.~Tkachev,
  ``Axion miniclusters and Bose stars,''
  Phys.\ Rev.\ Lett.\  {\bf 71}, 3051 (1993)
  [hep-ph/9303313].  

\bibitem{Namjoo:2017nia} 
  M.~H.~Namjoo, A.~H.~Guth and D.~I.~Kaiser,
  ``Relativistic Corrections to Nonrelativistic Effective Field Theories,''
  arXiv:1712.00445 [hep-ph].  

\bibitem{Raffelt:1996}
  G. G. Raffelt, \textit{Stars as Laboratories for Fundamental Physics},
  (University of Chicago Press, 1996), Chap. 14.
%
\bibitem{Kaplan:1985dv} 
  D.~B.~Kaplan,
  ``Opening the Axion Window,''
  Nucl.\ Phys.\ B {\bf 260}, 215 (1985).  
%
\bibitem{Srednicki:1985xd} 
  M.~Srednicki,
  ``Axion Couplings to Matter. 1. CP Conserving Parts,''
  Nucl.\ Phys.\ B {\bf 260}, 689 (1985).  
  
\bibitem{Kim:1979if} 
  J.~E.~Kim,
  ``Weak Interaction Singlet and Strong CP Invariance,''
  Phys.\ Rev.\ Lett.\  {\bf 43}, 103 (1979).
%
\bibitem{Shifman:1979if} 
  M.~A.~Shifman, A.~I.~Vainshtein and V.~I.~Zakharov,
  ``Can Confinement Ensure Natural CP Invariance of Strong Interactions?,''
  Nucl.\ Phys.\ B {\bf 166}, 493 (1980).
  
\bibitem{Dine:1981rt} 
  M.~Dine, W.~Fischler and M.~Srednicki,
  ``A Simple Solution to the Strong CP Problem with a Harmless Axion,''
  Phys.\ Lett.\  {\bf 104B}, 199 (1981).  
%
\bibitem{Zhitnitsky:1980tq} 
  A.~R.~Zhitnitsky,
  ``On Possible Suppression of the Axion Hadron Interactions. (In Russian),''
  Sov.\ J.\ Nucl.\ Phys.\  {\bf 31}, 260 (1980)
  [Yad.\ Fiz.\  {\bf 31}, 497 (1980)].  

\bibitem{McLachlan1947}
N.~W.~McLachlan,
\textit{Theory and Application of Mathieu Functions} (Oxford University Press, London, 1947), Chap. 6.
%
\bibitem{Erdelyi1955}
A.~Erd\'{e}lyi \textit{et al.}, 
\textit{Higher Transcendental Functions} (McGraw-Hill, New York, 1955),
Vol. 3, Chap. 16.
%
\bibitem{LandauLifshitz}
L.~D.~Landau and E.~M.~Lifshitz,
\textit{Mechanics: Course of Theoretical Physics} (Elsevier, Boston, 1976), pp. 80-84. 

\bibitem{Kolb:1993hw} 
  E.~W.~Kolb and I.~I.~Tkachev,
  ``Nonlinear axion dynamics and formation of cosmological pseudosolitons,''
  Phys.\ Rev.\ D {\bf 49}, 5040 (1994)
  [astro-ph/9311037].
  %
\bibitem{Visinelli:2017ooc} 
  L.~Visinelli, S.~Baum, J.~Redondo, K.~Freese and F.~Wilczek,
  ``Dilute and dense axion stars,''
  Phys.\ Lett.\ B {\bf 777}, 64 (2018)
  [arXiv:1710.08910 [astro-ph.CO]].  

\bibitem{Chavanis:2011zi} 
  P.~H.~Chavanis,
  ``Mass-radius relation of Newtonian self-gravitating Bose-Einstein condensates with short-range interactions: I. Analytical results,''
  Phys.\ Rev.\ D {\bf 84}, 043531 (2011)
  [arXiv:1103.2050 [astro-ph.CO]].
  %
  \bibitem{Chavanis:2011zm} 
  P.~H.~Chavanis and L.~Delfini,
  ``Mass-radius relation of Newtonian self-gravitating Bose-Einstein condensates with short-range interactions: II. Numerical results,''
  Phys.\ Rev.\ D {\bf 84}, 043532 (2011)
  [arXiv:1103.2054 [astro-ph.CO]].

\bibitem{Carlson:1994yqa}
E.~D.~Carlson and W.~D.~Garretson,
  ``Photon to pseudoscalar conversion in the interstellar medium,''
Phys.\ Lett.\ B {\bf 336}, 431 (1994).

\bibitem{Daido:2018dmu} 
  R.~Daido, F.~Takahashi and N.~Yokozaki,
  ``Enhanced axion�photon coupling in GUT with hidden photon,''
  Phys.\ Lett.\ B {\bf 780}, 538 (2018)
  [arXiv:1801.10344 [hep-ph]].
 %
\bibitem{Fan:2016rda} 
  J.~Fan,
  ``Ultralight Repulsive Dark Matter and BEC,''
  Phys.\ Dark Univ.\  {\bf 14}, 84 (2016)
  [arXiv:1603.06580 [hep-ph]].
  
\bibitem{Raffelt:1985nk} 
  G.~G.~Raffelt,
  ``Astrophysical Axion Bounds Diminished By Screening Effects,''
  Phys.\ Rev.\ D {\bf 33}, 897 (1986).

\bibitem{Hertzberg:2010yz} 
  M.~P.~Hertzberg,
  ``Quantum Radiation of Oscillons,''
  Phys.\ Rev.\ D {\bf 82}, 045022 (2010)
  [arXiv:1003.3459 [hep-th]].
  %
\bibitem{Kawasaki:2013awa} 
  M.~Kawasaki and M.~Yamada,
  ``Decay rates of Gaussian-type I-balls and Bose-enhancement effects in 3+1 dimensions,''
  JCAP {\bf 1402}, 001 (2014)
  [arXiv:1311.0985 [hep-ph]].
 

  
\end{thebibliography}
\end{document}